\documentclass[12pt]{article}

\makeatletter
\renewcommand\section{\@startsection {section}{1}{\z@}%
                                   {-3.5ex \@plus -1ex \@minus -.2ex}
                                   {2.3ex \@plus.2ex}%
                                   {\normalfont\large\bfseries}}
\renewcommand\subsection{\@startsection{subsection}{2}{\z@}%
                                     {-3.25ex\@plus -1ex \@minus -.2ex}%
                                     {1.5ex \@plus .2ex}%
                                     {\normalfont\bfseries}}
\makeatother

\usepackage{graphicx}
\usepackage{cite}

\newcommand{\beq}{\begin{equation}}
\newcommand{\eeq}{\end{equation}}
\newcommand{\ber}{\begin{array}}
\newcommand{\eer}{\end{array}}
\newcommand{\D}{{\cal D}}
\newcommand{\V}{{\cal V}}
\newcommand{\dtwo}{d^{\hspace{1pt}2}\hspace{-1pt}}
\newcommand{\del}{\partial}
\newcommand{\deln}{\partial_n}

\newcommand{\dsty}{\displaystyle}
\newcommand{\s}{\sigma}
\newcommand{\te}{\theta}

\newcommand{\de}{\delta}
\newcommand{\ds}{\dtwo\sigma}
\newcommand{\cnst}{\mbox{const}}
\newcommand{\tte}{\tilde\theta}

\newcommand{\eps}{\varepsilon}

\newcommand{\tlt}{\tilde t}

\def\IZ{\relax\ifmmode\mathchoice
{\hbox{\cmss Z\kern-.4em Z}}{\hbox{\cmss Z\kern-.4em Z}}
{\lower.9pt\hbox{\cmsss Z\kern-.4em Z}} {\lower1.2pt\hbox{\cmsss
Z\kern-.4em Z}}\else{\cmss Z\kern-.4em Z}\fi}
\def\IR{\relax{\rm I\kern-.18em R}}

\def\one{{\hbox{ 1\kern-.8mm l}}}

\newlength{\bredde}
\def\slash#1{\settowidth{\bredde}{$#1$}\ifmmode\,\raisebox{.15ex}{/}
\hspace*{-\bredde} #1\else$\,\raisebox{.15ex}{/}\hspace*{-\bredde}
#1$\fi}

\newsavebox{\zzzbar}
\sbox{\zzzbar}
  {\setlength{\unitlength}{0.9em}
  \begin{picture}(0.6,0.7)
  \thinlines
  \put(0,0){\line(1,0){0.6}}
  \put(0,0.75){\line(1,0){0.575}}
  \multiput(0,0)(0.0125,0.025){30}{\rule{0.3pt}{0.3pt}}
  \multiput(0.2,0)(0.0125,0.025){30}{\rule{0.3pt}{0.3pt}}
  \put(0,0.75){\line(0,-1){0.15}}
  \put(0.015,0.75){\line(0,-1){0.1}}
  \put(0.03,0.75){\line(0,-1){0.075}}
  \put(0.045,0.75){\line(0,-1){0.05}}
  \put(0.05,0.75){\line(0,-1){0.025}}
  \put(0.6,0){\line(0,1){0.15}}
  \put(0.585,0){\line(0,1){0.1}}
  \put(0.57,0){\line(0,1){0.075}}
  \put(0.555,0){\line(0,1){0.05}}
  \put(0.55,0){\line(0,1){0.025}}
  \end{picture}}

\newcommand{\ena}{\end{eqnarray}}
\newcommand{\beqa}{\begin{eqnarray}}
\newcommand{\eeqa}{\end{eqnarray}}
\newcommand{\bea}{\begin{eqnarray}}
\newcommand{\eea}{\end{eqnarray}}

\newcommand{\eq}[1]{(\ref{#1})}

\newcommand{\be}{\begin{equation}}
\newcommand{\ee}{\end{equation}}

\begin{document}

\begin{titlepage}
\begin{flushright}
hep-th/0609216, ITFA-2006-30
\end{flushright}
\vfill
\begin{center}
{\LARGE\bf D0-brane recoil revisited}    \\
\vskip 10mm
{\large Ben Craps,$^{a,b}$ Oleg Evnin$^{c,d}$ and Shin Nakamura$^{e,f}$}
\vskip 7mm
{\em $^{a}$ Theoretische Natuurkunde, Vrije Universiteit Brussel and\\
The International Solvay Institutes\\ Pleinlaan 2, B-1050 Brussels, Belgium\footnote{present address}}
\vskip 3mm 
{\em $^b$ Instituut voor Theoretische Fysica, Universiteit van Amsterdam, Valckenierstraat 65, 1018 XE Amsterdam, The Netherlands}
\vskip 3mm
{\em $^c$ California Institute of Technology 452-48, Pasadena, CA 91125, USA}  
\vskip 3mm
{\em $^d$ Department of Physics, Rostov State University, pr. Zorge 5,\\344090 Rostov-na-Donu, Russia}  
\vskip 3mm
{\em $^e$ Physics Department, Hanyang University, Seoul, 133-791, Korea}
\vskip 3mm
{\em $^f$ Center for Quantum Spacetime (CQUeST), Sogang University, Seoul, 121-742, Korea}
\vskip 3mm
{\small\noindent  {\tt Ben.Craps@vub.ac.be, eoe@caltech.edu, nakamura@hanyang.ac.kr}}
\end{center}
\vfill

\begin{center}
{\bf ABSTRACT}
\end{center}

\noindent 
One-loop string scattering amplitudes computed using the standard D0-brane conformal field theory (CFT) suffer from infrared divergences associated with recoil. A systematic framework to take recoil into account is the worldline formalism, where fixed boundary conditions are replaced by dynamical D0-brane worldlines. We show that, in the worldline formalism, the divergences that plague the CFT are automatically cancelled in a non-trivial way. The amplitudes derived in the worldline formalism can be reproduced by deforming the CFT with a specific ``recoil operator'', which is bilocal and different from the ones previously suggested in the literature. 
\vfill

\end{titlepage}

\section{Introduction}

There is nothing mysterious about the phenomenon of recoil. When a static compact object is hit, it starts moving in the direction of impact. This is indeed one of the simplest mechanical processes.
However, the amount of effort needed to formally accommodate the phenomenon of recoil does not always match its physical simplicity. One such example is the recoil of quantum solitons \cite{rajaraman}, heavy particle-like quantum descendants of localized classical static field configurations. The problem essentially comes from resorting to perturbative expansions in quantum field theory: due to the non-perturbatively large mass of the solitons, the recoil velocity is small, and the algebraic representation of recoil becomes distributed in a non-trivial way over the different orders of the perturbative expansion. Furthermore, the most na\"\i ve attempts to organize the perturbation theory in the presence of solitons are plagued by infrared divergences.

All of these difficulties have string-theoretic counterparts. If one tries to employ the standard worldsheet conformal field theory (CFT) construction of string scattering amplitudes in the background of a D0-brane, one ends up with infrared divergences in loop (e.g.\ annulus) diagrams. These divergences are associated with recoil. It is then imperative to find a way to re-organize the string perturbative expansion, if one is to be able to perform computations beyond leading order in the string coupling. 
The issue in string theory appears even more subtle than 
in field theory, since, as yet, a suitable non-perturbative 
formulation of string theory is unknown 
(whereas in field theory the conventional Lagrangian formulation 
is available).
One therefore does not have a more fundamental starting point for the investigations of recoil than the (infrared divergent) string perturbative expansion in the background of a static D0-brane.

Furthermore, one encounters difficulties specifying the initial and final states of the D0-brane, which is a necessity for implementing recoil. Indeed, open strings attached to the D0-brane give a highly singular description of its translational motion: on-shell massless modes in 0+1 dimensions necessarily have zero energy, so any finite velocity would require a way to deal with an infinite number of open strings. Even identifying the initial and final D0-brane states implicit in perturbative string scattering amplitudes is non-trivial. Classically, the standard Dirichlet boundary conditions describe a D0-brane with a well-defined position and zero velocity. Quantum mechanically (at non-zero string coupling), the D0-brane has to become delocalized in position or momentum space or both, and it is not a priori obvious which state is implicit in CFT computations. 

Various strategies have been proposed for implementing recoil via extensions of the worldsheet CFT \cite{tafjord, fischler, hirano-kazama, shin, worldline}. One class of proposals proceed along the lines of the Fischler-Susskind mechanism \cite{fischler-susskind,polchinski-fischler-susskind} and construct a deformation of the standard D0-brane CFT (referred to as ``recoil operator'') that would represent a recoiling D0-brane \cite{tafjord,fischler}. Unfortunately, the condition that the recoil operator should cancel the annulus divergence does not appear to fix it uniquely (in particular, different recoil operators have been proposed in \cite{tafjord} and \cite{fischler}). One can furthermore show that, if one cancels the annular divergence using the recoil operators of \cite{tafjord} or \cite{fischler}, the resulting finite amplitudes display pathological features (see appendix A, and the discussion in section 3).

In the second class of proposals, one introduces fully dynamical worldlines for the D0-brane, and integrates over all the possible trajectories \cite{hirano-kazama,worldline}. This approach creates within the theory an explicit dynamical variable describing the translational motion of the D0-brane. Hence explicitly specifying the initial and final states of its translational motion no
longer poses a problem. The worldline formalism was first introduced in \cite{hirano-kazama}, but it was not until \cite{worldline} that adequate techniques were developed to perform computations with fully dynamical D0-brane worldlines.

In the present paper, we use the formalism of \cite{worldline} to compute string scattering amplitudes in the presence of a D0-brane. The main result is that, in the worldline formalism, the annulus divergences that plagued standard CFT computations are automatically cancelled in a technically non-trivial way by divergent disk contributions. The disk contributions can be reproduced by deforming the CFT with a specific ``recoil operator'', which is bilocal and different from the ones previously suggested in the literature. This way, we make contact between the two main strategies to implement D0-brane recoil in perturbative string theory: with current technology, the worldline formalism appears to be the more systematic way to compute scattering amplitudes, but the results are consistent with introducing a recoil operator in the CFT (although the precise form of the recoil operator would have been hard to derive purely within CFT).

A complementary perspective on D0-brane recoil is arrived at using low-energy effective field theory, in particular supergravity coupled to the Dirac-Born-Infeld (DBI) action of a D0-brane \cite{shin}. We use a DBI analysis to provide evidence that the infrared divergent annulus contribution to scattering amplitudes in the presence of a static D0-brane can be combined with divergent contributions from worldsheets with more holes, in such a way that the sum of all contributions vanishes if non-zero momentum is transferred to the D0-brane. In the DBI analysis, we show that the vanishing of the resummed amplitude is due to momentum conservation: the amplitude has to be zero if momentum is not conserved among the closed strings and D0-brane recoil is not taken into account. In the course of the analysis, we provide evidence that the quantum state implicitly selected by the standard D0-brane CFT is sharply localized in momentum space (and has a singular normalization factor).   

\section{The annular divergence}

The annular divergence in the background of a static D0-brane is a principal point of departure for investigations of recoil, since it is this divergence that signals the breakdown of the standard CFT description of a recoiling D0-brane. For a discussion of the annulus amplitude in the context of more general D-branes, we refer to \cite{local}.

For solitons in field theory, infrared divergences in loop diagrams come from large distance propagation of the zero modes corresponding to shifting the entire topological defect. In string theory, such large distance propagation corresponds to the annulus developing a long, thin strip. Divergences from degenerating Riemann surfaces can be analyzed using Polchinski's plumbing fixture construction \cite{polchinski-fischler-susskind}, which relates the divergences to amplitudes evaluated on a lower genus Riemann surface. In particular, the annulus amplitude with an insertion of vertex operators $V^{(1)},\cdots,V^{(n)}$ (in the interior) can be expressed through disk amplitudes with additional operator insertions at the boundary:
\beq
\left\langle V^{(1)}\cdots V^{(n)}\right>_{annulus}=\sum\limits_\alpha \int \frac{dq}q\, q^{h_\alpha-1} \int d\te d\te' \left\langle V_\alpha(\te)V_\alpha(\te')V^{(1)}\cdots V^{(n)}\right>_{D_2},
\label{plumbing}
\eeq
where the summation extends over a compete set of local operators $V_\alpha(\te)$ with conformal weights $h_\alpha$, and $q$ is the gluing parameter, which can be related to the annular modulus. ($\te$ parametrizes the boundary of the disk.) The divergence in the integral over $q$ coming from the region $q\approx0$ (i.e., from an annulus developing a thin strip) will be dominated by the terms with the smallest possible $h_\alpha$. 

Neglecting the tachyon divergence, which is a pathology peculiar to the case of the bosonic string, we consider (in close relation to the investigations of \cite{fischler}) the following operators with conformal weights $h=1+\alpha'\omega^2$:
\be
V^i(\te)=\ \del_n X^i(\te)\exp\left[i\omega X^0(\te)\right]\ .
\ee
These operators correspond to massless open string states (representing translations of the D0-brane in the $i$'th Dirichlet direction). For small values of $q$ (which is the region we are interested in), only small values of $\omega$ will contribute to the integral. Hence, the annular divergence takes the following form:
\bea
\left\langle V^{(1)}\cdots V^{(n)}\right>_{annulus}^{(div)}&\sim&\int\limits_0^1 dq\int\limits_{-\infty}^\infty  d\omega\, q^{-1+\alpha'\omega^2}\int d\te d\te' \left\langle V^i(\te,\omega)V^i(\te',\omega)V^{(1)}\cdots V^{(n)}\right>_{D_2}\nonumber\\
&\sim&\int\limits_0^1 dq\int\limits_{-\infty}^\infty  d\omega\, q^{-1+\alpha'\omega^2}\int d\te d\te' \left\langle V^i(\te,0)V^i(\te',0)V^{(1)}\cdots V^{(n)}\right>_{D_2}\nonumber\\
&\sim& P^2\left\langle V^{(1)}\cdots V^{(n)}\right>_{D_2}\int\limits_0^1 dq\int\limits_{-\infty}^\infty d\omega\, q^{-1+\alpha'\omega^2}\nonumber\\
&\sim&P^2\left\langle V^{(1)}\cdots V^{(n)}\right>_{D_2}\int\limits_0^1 {dq\over q\,(-\log q)^{1/2}}\ ,
\label{annulus}
\eea
where in the transition from the second to the third line we have taken into account the fact that the operator $\int \del_n X^i(\te)d\te$ merely shifts the position of the D0-brane; inserting it into any amplitude amounts to multiplication by the total (Dirichlet) momentum $P$ transferred by the closed strings to the D0-brane during scattering.

Introducing a cut-off $\eps$ on the lower bound of the integral (\ref{annulus}) reveals a  $\sqrt{|\log\eps|}$ divergence\footnote{Let us note in passing that, for the case of scattering off a D1-brane, one encounters a $\log|\log\eps|$ divergence instead. Just as recoil provides a physical interpretation for the annular divergence in the background of a D0-brane, the divergence in the background of a D1-brane must be given a clear intuitive explanation. The associated phenomenon, which we call ``local recoil'', can indeed be identified, and it has been described in a separate publication \cite{local}.}, which is indicative of recoil:
\beq
\left\langle V^{(1)}\cdots V^{(n)}\right>_{annulus}^{(div)}\sim P^2\left<V^{(1)}\cdots V^{(n)}\right>_{D_2}\sqrt{|\log\eps|}.
\label{annulusD0}
\eeq
The overall normalization in this expression can be fixed, for example,
through an appeal to the DBI formalism; the derivations are given
in appendix C. The resulting expression for the annular divergence is
\beq
\left\langle V^{(1)}\cdots V^{(n)}\right>_{annulus}^{(div)}=-\frac{P^2}{2M}\,\sqrt\frac{\alpha'}{\pi}\,\left< V^{(1)}\cdots V^{(n)}\right>_{D_2}\sqrt{|\log\eps|},
\label{appendixC}
\eeq
where $M$ is the mass of the D0-brane.

\section{The final state of the recoiling D0-brane} 

It is a common intuition (also building upon the results on solitons in quantum field theory) that the annular divergence in the background of a static D0-brane is caused by an improper account of recoil. Indeed, propagating closed strings witness a recoiling D0-brane, which cannot be viewed as a ``small perturbation'' of the static D0-brane background one tries to expand around in standard perturbative string theory. One should therefore hope that implementing the background of a recoiling D0-brane within the formalism would eliminate the divergence. In other words, we should be trying to construct closed string scattering amplitudes for which the velocity of the D0-brane is different in the asymptotic past and the asymptotic future. Unfortunately, as we discussed in the introduction, it is not possible to implement such a program in a straightforward way. Indeed, how would one construct a state corresponding to a moving D0-brane in the standard D0-brane CFT?

It is commonly mentioned that the dynamical states of D-branes are represented in the formalism of perturbative string theory as the massless scalar vibrational states of the open strings attached
to the D-brane worldvolume. For the purpose of evaluating the S-matrix, such massless scalar vibrational states should be represented by their vertex operators:
\be
\int d\te :\del_n X^i(\te) \exp\left[ik_{\mu}X^{\mu}(\te)\right]:
\ee
(where the index $i$ runs over the Dirichlet directions and the index $\mu$ over the Neumann directions). Indeed, for (non-compact) higher-dimensional D-branes, such massless states of open strings attached to the D-brane can be identified with one-particle states of the worldvolume fields corresponding to the deformations of the D-brane. This is quite intuitive: the open strings move at the speed of light in the Neumann directions, and so do the excitations of the worldvolume. Furthermore, should one be willing to specify the initial and final vibrational states of a higher-dimensional brane in a scattering process involving, say, closed strings, this can be immediately accomplished by including the appropriate open strings into the initial and final states of the string theory S-matrix. 

One must, however, realize that a significant subtlety is encountered if one attempts to extend this picture to the case of D0-branes. Indeed, the only dynamical degree of freedom of the D0-brane is
the translational mode (i.e.\ the spatial coordinate of the D0-brane). The quantum-mechanical spectrum of such mode is well known to be comprised of momentum eigenstates,
in particular, it is a continuous spectrum. On the other hand, as mentioned in the introduction, the on-shell energy of the ``massless scalar'' open string states attached to the D0-brane
is exactly zero. 
The energy spectrum of the translational mode generated by such ``massless scalar'' open strings will be singular, with discrete, infinitesimally spaced
levels.
While heuristically one can try to think of the non-zero momentum states as containing an infinite number of open strings, the practical value of this picture is limited, unless one is able to specify a recipe for how the open strings should be used to describe moving D0-branes in the standard D0-brane CFT.

Using open strings to describe the translational mode of the D0-brane would be analogous to using the zero frequency limit of the harmonic oscillator to describe a free particle (the quanta of the harmonic oscillator being analogous to the open strings). Note that a similar problem emerges for field theory solitons if one does not treat the translational mode carefully. On the other hand, an explicit introduction of the translational mode for the solitons (as, for example, per Christ-Lee method \cite{Christ:1975wt}) resembles the worldline formalism for D0-branes. For a pedagogical discussion of the translational motion of solitons, see \cite{rajaraman}.

What are the alternatives to the singular description of the translational motion of the D0-brane by means of open strings? In analogy to the techniques used for field theory solitons, one can introduce the translational degree of freedom for the D0-brane explicitly (as a dynamical trajectory). The (singular) initial and final state massless open strings are then absent from the physical amplitudes by construction. This is the worldline formalism that will be our primary subject in the next section. As we shall see then, the amplitudes constructed within this framework are free of infrared pathologies.

There is a direct analogy between the approach we have just described and the standard treatment of D-instantons \cite{combinatorics}. Namely, in the latter case, one does not attempt to describe the translational zero-modes of the D-instantons in terms of open strings. Instead, one explicitly introduces the collective coordinate (i.e., the position of the D-instanton) and integrates over it. The resulting amplitudes are free of infrared pathologies. The worldline formalism is a direct analog of this procedure for the case of D0-branes.

Before we proceed with a detailed construction of the worldline formalism and evaluation of the scattering amplitudes, we would like to discuss briefly the approach to recoil advocated in \cite{tafjord}, since the recoil operator proposed in that paper bears some similarities to the one we will find using the worldline formalism. A more detailed critical review of the early literature \cite{tafjord, fischler, hirano-kazama} is presented in appendix A.

It is a common paradigm in string theory \cite{fischler-susskind, polchinski-fischler-susskind} that, when infrared divergences are present in loop diagrams, one must add to the worldsheet action a term (the ``Fischler-Susskind operator'') that explicitly involves the string coupling. The Fischler-Susskind operator should be such that the higher-loop divergences of the original theory are cancelled by lower-loop divergences with insertions of the Fischler-Susskind operator. For the case of D0-brane recoil, one would say that the modified worldsheet action represents a recoiling (rather than static) D0-brane, and refer to the corresponding deformation of the string action as the ``recoil operator''.

In \cite{tafjord}, it is suggested to cancel the annulus divergence by introducing the following recoil operator:
\be\label{PTop}
V_{PT}\sim v^i\int d\te\,\del_n X^i(\te)\, X^0(\te)\,\Theta(X^0(\te))
\ee
(with $\Theta(t)$ being the step function, and $v^i$ the final velocity of the D0-brane, which is of the order of the string coupling). The hope is then that the background correction due to $V_{PT}$ will introduce a divergence on the disk
\be
\langle V_{PT} V_1(\s_1)\cdots V_n(\s_n)\rangle_{D_2}
\ee
which will, in turn, cancel the annulus divergence (\ref{annulus}). The physical interpretation of the background modification by $V_{PT}$ is that the D0-brane abruptly starts moving with the appropriate recoil velocity $v^i$ at the moment $X^0=0$. Note that the classical trajectory of the D0-brane implied in $V_{PT}$ is given as $v^i\,X^0\,\Theta(X^0)$.

The operator $V_{PT}$ has indeed the right structure to cancel the annulus divergence (\ref{annulus}). However, $V_{PT}$ implies that the recoil happens at a given moment of time ($X^0=0$), and 
as discussed in appendix A, the physical finite part of the amplitude does depend on which moment one chooses in the recoil operator. Since there is no physically meaningful ``moment of recoil'' for quantum particles with well-defined energies, such an ambiguity is unacceptable.

Our main tool in this present investigation of the D0-brane recoil will be the worldline formalism (rather than explicit deformations of the D0-brane CFT). Nevertheless, once the scattering amplitudes have been computed within the worldline formalism, it will be possible to see that the results can, in fact, be reproduced by deforming the CFT with an appropriate recoil operator.
This operator will be reminiscent of the operator \eq{PTop}, and it will have a fairly clear heuristic interpretation, but we should emphasize that it is distinct from \eq{PTop} (for instance, it is bilocal rather than local) and it will not imply that the D0-brane moves along a classical trajectory.

\section{Worldline approach to D0-brane recoil}

We now proceed to construct the dynamical worldline description of D0-branes. This approach has been originally proposed in \cite{hirano-kazama} and considerably strengthened and reorganized in \cite{worldline, eoe}. In the present exposition, we first give a more pedagogical account of the formalism of \cite{worldline, eoe}, focusing on the aspects of the worldline derivations directly relevant to the problem of recoil. Then we extend the computations of \cite{worldline, eoe} to demonstrate the cancellation of divergences and derive a recoil operator that allows to reproduce our results in conformal field theory.

\subsection{Quantization of D0-brane worldlines}

It is the principal objective of the worldline formalism to give an adequate account of the translational motion of the D0-brane. To this end, one introduces its coordinates explicitly and integrates over all the possible worldlines $f^\mu(t)$, with $t$ being the proper time. The boundaries of the string worldsheet are restricted to the D0-brane worldline, and the emission
of closed strings is described by insertions of the closed string vertex operators in the interior of the worldsheet. Our main object of interest is the amplitude for a D0-brane to move from the point $x_1^\mu$ to the point $x_2^\mu$ while absorbing/emitting $m$ closed strings carrying momenta $k_1$ to $k_m$:
\beq
\ber{l}
\dsty G(x_1,x_2|\,k_1,\cdots,k_m)=\sum\frac{\left(g_{st}\right)^\chi}{V_\chi} {\int[\D f]}_{\mbox{\small diff}} \D t\,\D X\,\de\left(X_\mu(\te)-f_\mu(t(\te))\right)\vspace{2mm}\\
\dsty\hspace{4.5cm}\times\exp\left[-S_{D}(f)-S_{st}(X)\right] \prod\limits_{a=1}^m\left\{g_{st}\V_a(k_a)\right\},
\eer
\label{master}
\eeq
where $S_D$ is the action for the D0-brane to be discussed below, $S_{st}$ is the standard conformal gauge action
\be
S_{st}=\frac1{4\pi\alpha'}\int\ds\nabla X_\mu\nabla X^\mu,
\ee
the integration with respect to $f_\mu$ extends over all the inequivalent (unrelated by diffeomorphisms) curves starting at $x_1$ and ending at $x_2$, the boundary of the worldsheet is parametrized by $\te$, and $t(\te)$ describes how this boundary is mapped onto the D0-brane worldline. The sum is over all the topologies of the worldsheets (not necessarily connected, but without any disconnected vacuum parts) and $\chi$ is the Euler number. $V_\chi$ is the conformal Killing volume (the negative regularized value of \cite{volume} should be used for the disk). The fully integrated form of the vertex operators is implied. We work in Euclidean space-time, keeping in mind a subsequent analytic continuation to Minkowski signature. The integration over moduli
of the worldsheet is suppressed, since we shall be mostly working with worldsheets of disk topology. The scattering amplitude can be deduced from (\ref{master}) by means of the standard reduction formula:
\be
\left<p_1|p_2\right>_{k_n}=\lim\limits_{p_1^2,p_2^2\to -M^2} \left(p_1^2+M^2\right)\left(p_2^2+M^2\right)\int dx_1dx_2 e^{ip_1x_1} e^{ip_2x_2}G(x_1,x_2|\,k_1,\cdots,k_m), 
\label{reduct}
\ee
where $M$ is the D0-brane mass.

Weyl invariance appears to be a rather subtle issue in (\ref{master}). On physical grounds, one would believe that making the D-branes fully dynamical reinforces the consistency of the amplitudes, much in the same way as respecting the supergravity equations of motion makes the non-linear $\s$-models consistent. This issue is, of course, intimately related to the cancellation of divergences in the worldsheet integration, which is so essential to the implementation of recoil. This cancellation of divergences will be the central theme of our derivations within the worldline formalism, and we shall show
(technically, to next-to-leading order) that the theory is indeed divergence-free.

Of course, whether or not the integration over the D0-brane worldlines reinforces the consistency of the string amplitudes depends crucially on the choice of the D0-brane worldline action. It appears to be a fairly general principle \cite{fradkin} that the value of the effective action for a background that couples to strings is given by (minus) the sum of all connected vacuum string graphs evaluated in this background. Thus, very much in the spirit of \cite{combinatorics}:
\begin{equation}
S_D[f]=\sum\limits_{\mbox{connected}}{\frac{\left(g_{st}\right)^\chi}{-V_\chi}\int\D t\,\D X \,\de\left(X_\mu(\te)-f_\mu(t(\te))\right)\exp\left[-S_{st}(X)\right]}.
\label{SD}
\end{equation}
Again, the negative regularized value of the conformal Killing volume should be used for the disk \cite{volume}. The exponentiation of the action in the path integral can be seen as a result of summing up the disconnected graphs containing vacuum parts \cite{combinatorics}. It can be shown\footnote{Essentially, $M$ is the constant produced by the integration over the non-zero modes of $t(\te)$, and the factor of $T$ comes from the integration over the zero mode of $t(\te)$.} that, for nearly straight worldlines, the above action reduces to the na\"\i ve point-particle result
$MT$ (with $T$ being the length of the worldline). For curved worldlines, (\ref{SD}) would take into account the backreaction from the spacetime fields excited by the accelerating D0-brane. One must realize, however, that, for our present purposes (i.e., for investigations of D0-brane recoil at next-to-leading order in the string coupling), it should suffice to set $S_D=MT$. Indeed, one can generically write an expansion of $S_D$ around a straight worldline:
\be
\ber{l}
\dsty S_D[f]=MT+\int dt_1\,dt_2\, C_2(t_1,t_2) f^i(t_1)f_i(t_2)\vspace{2mm}\\
\dsty\hspace{2cm}+\int dt_1\,dt_2\,dt_3\,dt_4\, C_4(t_1,t_2,t_3,t_4) f^i(t_1)f_i(t_2)f^j(t_3)f_j(t_4)+\cdots\ ,
\eer
\ee
where the indices $i$ and $j$ run over the Dirichlet directions. Note, that only even powers of $f$ can be present by Lorentz invariance. Furthermore, due to translational invariance, all the entries of $f(t)$ can be replaced, say, by $f(t)-f(0)$ (i.e., by a difference in the Dirichlet position between two points). But, in a recoil process, due to the non-perturbatively large mass of the D0-brane, all the velocities, and hence all the position differences are of order $g_{st}$. Therefore, the above expansion implies that
\be
S_D[f]=MT+O(g_{st}^2).
\ee
But, since we intend to work at the order $g_{st}$, rather than $g_{st}^2$, we can set\footnote{Somewhat surprisingly, the derivations within the worldline formalism can be carried out with a good deal of success even if the full action (\ref{SD}) is employed. We shall not need such exact derivations for our present purposes, and will refer the interested readers to the considerations of \cite{worldline} and \cite{eoe}.}
\beq
S_D[f]=MT
\label{MT}
\eeq
for the rest of our present considerations.

Using the transformation properties of the vertex operators under the target space translations, it is easy to see that
\be
G(x_1,x_2|\,k_n)=\exp\left[\frac{i}2(x_1^\mu+x_2^\mu)\sum k_n\right] G\left(\left.\frac{x_1-x_2}2,-\frac{x_1-x_2}2\right|\,k_n\right).
\ee
The first term here merely provides for the momentum conservation $\de$-function in the Fourier transform, and (\ref{reduct}) can be rewritten as%
\footnote{Note that, in our conventions, for the incoming particles, $p$ (or $k$) is the momentum, and, for the outgoing particles, it is minus the momentum. Hence, the energy-momentum conservation has the form $p_1+p_2+\sum k_n=0$.
}
\beq
\ber{l}
\dsty\left<p_2|p_1\right>_{k_n}=(2\pi)^{26}\de(p_1+p_2+\sum k_n)\vspace{2mm}\\
\dsty\hspace{2cm}\times\lim\limits_{p_1^2,p_2^2\to -M^2} \left(p_1^2+M^2\right)\left(p_2^2+M^2\right)\int dx \exp\left[\frac{i}2(p_1-p_2)x\right] G\left(\left.\frac{x}2,-\frac{x}2\right|k_n\right). 
\eer
\label{reduct1}
\eeq
We shall work with this representation in our subsequent calculation of the amplitude.

\subsection{The Gaussian integration}

{}From (\ref{master}), it is easy to see that the integration over $X$ is Gaussian and can be performed exactly. We will thus be able to recast the formalism into a (0+1)-dimensional form. The Gaussian integration we have to perform is closely related to the derivations in \cite{fradkin} and can be implemented by applying the formula%
\footnote{The derivation of this basic yet important formula together with a few underlying subtleties is described in appendices A and B of \cite{eoe}. One can easily understand the general structure after shifting the integration variable $X(\s)$ by a solution of the Laplace equation $\bar X(\s)$ satisfying the boundary condition $\bar X(\te)=\xi(\te)$. The second and third terms in the exponent of (\ref{gaussian}) originate from the change of variables, whereas the first term arises from the remaining Dirichlet Gaussian integration. (Note that the boundary-to-boundary and boundary-to-interior propagators satisfy a number of identities described in \cite{eoe}. In particular, there are identities relating the Neumann and Dirichlet Green functions.)
}
\beq
\ber{l}
\dsty\int\D X\,\de\left(X(\te)-\xi(\te)\right)\, \exp\left[\int \ds\left(-\frac1{4\pi\alpha'}\nabla X\nabla X+iJX\right)\right]\vspace{4mm}\\
\dsty\hspace{2cm}=\exp\left[-\pi\alpha'\int J(\s)D(\s,\s')J(\s')\ds\ds'-i\int\xi(\te)\deln D(\te,\s')J(\s')d\te\ds'\right.\vspace{2mm}\\
\dsty\hspace{5cm}\left. +\frac1{4\pi\alpha'}\int \xi(\te)\del_n\del_{n'}D(\te,\te')\xi(\te')d\te d\te'\right].
\eer
\label{gaussian}
\eeq
Here, $D$ is the Dirichlet Green function of the Laplace operator, $\Delta D(\s,\s')=-\de(\s-\s')$, and $\del_n$ denotes the normal derivative evaluated at the boundary (which is parametrized
by $\theta$). It is convenient to consider
\beq
\ber{l}
\dsty G\left(\left.\frac{x}2,-\frac{x}2\right|J\right)=\sum\frac{\left(g_{st}\right)^\chi}{V_\chi} {\int[\D f]}_{\mbox{\small diff}} \D t\,\D X\, e^{-MT}\,\de\left(X_\mu(\te)-f_\mu(t(\te))\right)\vspace{2mm}\\
\dsty\hspace{4.5cm}\times\exp\left[-S_{st}(X)+i\int \ds J_\mu X^\mu\right]
\eer
\label{master2}
\eeq
instead of (\ref{master}). Indeed, differentiating%
\footnote{More specifically, the vertex operators are polynomials of $\del X^\mu/\del\s^\alpha$ times $e^{ik\cdot X}$. One can simulate the insertion of $\del X^\mu/\del\s^\alpha$ by the functional differentiation
\be
\frac{\del}{\del\s^\alpha}\,\frac{\de}{\de J_{\mu}(\s)}.
\ee
Setting $J=\sum k_n\de(\s-\s_n)$ at the end of the computation will take care of the factor of $e^{ik\cdot X}$.
} 
with respect to the source $J$ and setting it to $\sum k_n\de(\s-\s_n)$ allows us to reproduce the amplitude for an arbitrary vertex operator insertion. Performing the integration in (\ref{master2}) by means of (\ref{gaussian}) yields
\beq
\ber{l}
\dsty G\left(\left.\frac{x}2,-\frac{x}2\right|J\right)=\sum\frac{\left(g_{st}\right)^\chi}{V_\chi} \exp\left[-\pi\alpha'\int J^\mu(\s)D(\s,\s')J_\mu(\s')\ds\ds'\right]\vspace{2.5mm}\\
\dsty\hspace{2cm}{\int[\D f]}_{\mbox{\small diff}}\,\D t\, e^{-MT}\, \exp\left[-i\int f^\mu(t(\te))\deln D(\te,\s')J_\mu(\s')d\te\ds'\right]\vspace{4mm}\\
\dsty\hspace{5cm}\exp\left[ \frac1{4\pi\alpha'}\int f^\mu(t(\te))\del_n\del_{n'}D(\te,\te')f_\mu(t(\te')) d\te d\te'\right].
\eer
\label{0+1}
\eeq
It should be noted that $D(\s,\s')$, $\chi$ and $V_\chi$ depend on the topology of the diagram corresponding to each particular term in the sum. For diagrams with disconnected parts, $D(\s,\s')$ is block diagonal in the sense that it vanishes whenever the two arguments belong to different disconnected components.

The path integral in (\ref{0+1}) may seem rather cumbersome, as one of the functions to be integrated over appears in the argument of the other one. Nevertheless, the integration over $f^\mu(t)$ can be performed exactly by means of a technique very similar to the treatment of the free point particle in \cite{polyakov}. We first rewrite the measure on reparametrization equivalence classes of $f^\mu(t)$ as
\be
[\D f]_{\mbox{\small diff}}=\D f\,\de\left[\dot{f}^2-1\right]=\D f \int\D z \exp\left[-\int\limits_0^Tz(\dot{f}^2-1)dt\right], 
\ee
where $\de[\dot{f}^2-1]$ is a product of $\de$-functions at every point (reinforcing $t$ to be the proper time), and, for each $t$, the integration over $z(t)$ is along a contour going from $c-i\infty$ to $c+i\infty$ in the complex $z$ plane, with $c$ being an arbitrary (positive) constant. This contour can of course be deformed, an opportunity implicit in our subsequent application of the saddle point method. If we now introduce%
\footnote{These shorthands have been first proposed in \cite{worldline}. They may require a certain amount of time to get accustomed to, but ultimately prove very convenient in handling the formalism. The general algebraic structure here is as follows: whenever there are two functions $f(\te)$ and $t(\te)$ at our disposal, we can introduce
\be
\tilde f(t)\equiv \int d\te\,\de(t-t(\te))\,f(\te).
\ee
In particular,
\be
\int \tilde f(t)\,dt = \int f(\te)\,d\te
\ee
and, if $f(\te)=g(t(\te))$, then $\tilde f(t)=g(t)$. The purpose of such transformations is to remove $t(\te)$ from the arguments of the functions appearing in the worldline path integral, which is an important pre-requisite for explicitly performing the Gaussian path integration.
}
\be
\ber{c}
\dsty{\cal N}(t,t')=-\int d\te d\te'\,\del_n\del_{n'}D(\te,\te')\, \de\left(t-t(\te)\right)\de\left(t'-t(\te')\right);\vspace{3mm}\\
\dsty d(t,\s)=\int d\te\,\deln D(\te,\s)\,\de\left(t-t(\te)\right),
\eer
\ee
the $f$-integration in (\ref{0+1}) can be recast into a manifestly Gaussian form:
\be
\ber{l}
\dsty G\left(\left.\frac{x}2,-\frac{x}2\right|J\right)=\sum\frac{\left(g_{st}\right)^\chi}{V_\chi}
\exp\left[-\pi\alpha'\int J^\mu(\s)D(\s,\s')J_\mu(\s')\ds\ds'\right]\vspace{2.5mm}\\
\dsty\hspace{2cm}\int\D t\,\D z\,\D f \, e^{-MT}\,\exp\left[-\int z(\dot{f}^2-1)dt\right]
\exp\left[-i\int f^\mu(t)d(t,\s)J_\mu(\s)dt\ds\right]\vspace{5mm}\\
\dsty\hspace{5cm}
\exp\left[-\frac1{4\pi\alpha'}\int f^\mu(t){\cal N}(t,t')f_\mu(t')dt dt'\right].
\eer
\ee

It is convenient to change to integration over $\dot{f}$ by means of the relations%
\footnote{Note that the question of what measure one should choose for the integration over $T$ is a priori subtle and it has been given an extensive treatment in \cite{polyakov}. The na\"\i ve measure $dT$ (rather than $\mu(T)dT$) is correct in our case. In particular, with this measure, the integration over $T$ gives the correct pole structure necessary for the reduction formula, whereas a different choice would not have produced kinematically acceptable momentum dependences of the (off-shell) amplitudes.
}
\beq
\D f = dT\D\dot{f}\,\de\left(\int\limits_0^T\dot{f}^\mu dt+x^\mu\right),\qquad f^\mu(t)=\frac{x^\mu}2+\int\limits_0^t\dot{f}^\mu dt.
\label{diffpath}
\eeq
It is also important to keep in mind the various properties of the functions ${\cal N}(t,t')$ and $d(t,\s)$. In particular, the following relations for the Dirichlet Green function (which are intimately related to the two-dimensional Gauss law)
\be
\int d\te\,\del_n\del_{n'}D(\te,\te')=0,\qquad\int d\te\,\deln D(\te,\s')=-1
\ee
imply that
\beq
\int\limits_0^T dt\,{\cal N}(t,t')=0,\qquad\int\limits_0^T dt\,d(t,\s)=-1.
\label{Nd}
\eeq
If we also perform the Fourier transformation (\ref{reduct1}) and substitute the Fourier representation of the $\de$-function in (\ref{diffpath}), we arrive at the expression
\be
\ber{l}
\dsty G(p_1,p_2|J)=\int dx \exp\left[\frac{i}2(p_1-p_2)x\right] G\left(\left.\frac{x}2,-\frac{x}2\right|J\right)\vspace{3mm}\\
\dsty\hspace{1.7cm}=\sum\frac{\left(g_{st}\right)^\chi}{V_\chi}\exp\left[-\pi\alpha'\int J^\mu(\s)D(\s,\s')J_\mu(\s')\ds\ds'\right]\vspace{3mm}\\
\dsty\hspace{5mm}\int dx\,dT \D t\,\D z\,\,e^{\int z\,dt}\, e^{-MT}\int\D\dot{f} \,\, e^{-\int z\dot{f}^2dt}\,\exp\left[ix\left(\frac{p_1-p_2}2+\frac12\int Jd\s\right)\right]
\delta\left(\int\limits_0^T\dot{f}^\mu dt+x^\mu\right)
\vspace{3mm}\\
\dsty\hspace{4cm}
\exp\left[-i\int dt\ds\left(\int\limits_0^t \dot{f}_\mu(\tilde t)d\tilde t\right)d(t,\s)J^\mu(\s)\right]\vspace{5mm}\\
\dsty\hspace{5cm}
\exp\left[-\frac1{4\pi\alpha'}\int dt dt'\left(\int\limits_0^t \dot{f}_\mu(\tilde t)d\tilde t\right){\cal N}(t,t')\left(\int\limits_0^{t'} \dot{f}_\mu(\tilde t')d\tilde t'\right)\right].
\eer
\ee
We now perform the integration over $x$, taking into account that $\int J d\s=-p_1-p_2$, and also switching the order of integrations in the exponents of the last two lines, while keeping in mind the relations (\ref{Nd}). The result takes the form
\be
\ber{l}
\dsty G(p_1,p_2|J)=\sum\frac{\left(g_{st}\right)^\chi}{V_\chi}
\exp\left[-\pi\alpha'\int J^\mu(\s)D(\s,\s')J_\mu(\s')\ds\ds'\right]\vspace{3mm}\\
\dsty\hspace{.5cm}\int dT \D t\,\D z\,\,e^{\int z\,dt}\, e^{-MT}\int\D\dot{f} \,
\exp\left[-i\int \dot{f}_\mu(t)\left(p_1^\mu-\int\limits_0^t
d(\tilde t,\s)J^\mu(\s)d\tilde t\ds\right)dt\right]\vspace{3mm}\\
\dsty\hspace{5cm}
\exp\left[-\int \dot{f}^\mu(t){\cal B}(t,t')\dot{f}_\mu(t')dt dt'\right],
\eer
\ee
where we have introduced
\be
{\cal B}(t,t')=z(t)\,\de(t-t')+\frac1{4\pi\alpha'}\int\limits_0^t d\tilde t \int\limits_0^{t'} d\tilde t'\,{\cal N}(\tilde t,\tilde t').
\ee
At this point, the Gaussian integration becomes completely straightforward and yields
\beq
\ber{l}
\dsty G(p_1,p_2|J)=\sum\frac{\left(g_{st}\right)^\chi}{V_\chi} \exp\left[-\pi\alpha'\int J^\mu(\s)D(\s,\s')J_\mu(\s')\ds\ds'\right]\vspace{3mm}\\
\dsty\hspace{4cm}\int dT \D t(\te)\,\D z(t)\,\det[{\cal B}]^{-D/2}\, e^{-MT}\,\exp\left[\int z\,dt\right]\vspace{2mm}\\
\dsty\exp\left[-\frac14\int \Big(p_1^\mu-\int\limits_0^t d(\tilde t,\s)J^\mu(\s)d\tilde t\ds\Big) {\cal B}^{-1}(t,t')\Big(p_{1\mu}-\int\limits_0^{t'} d(\tilde t',\s')J_\mu(\s')d\tilde t'\ds'\Big)
dtdt'\right].
\eer
\label{zt}
\eeq
From this representation, it is apparent that the endpoints of the $z$ integration contour can be moved towards $-\infty$. The contour itself cannot shrink to $-\infty$, however, on account of the singularities of $(\det{\cal B})^{-D/2}$. Should there be a discontinuity in $t(\te)$, these singularities move towards $-\infty$ allowing the contour to be deformed%
\footnote{
To make this more specific, one can consider the quadratic form defined by ${\cal B}(t,t')$:
\be
\int dt\,dt' \varphi(t){\cal B}(t,t')\varphi(t')=\int z\varphi^2 dt -\frac1{4\pi\alpha'}\int d\te d\te'\del_n\del_{n'}D(\te,\te')\int\limits_0^{t(\te)} \varphi(t)dt\int\limits_0^{t(\te')}\varphi(t') dt',
\label{Bquad}
\ee
where $\varphi(t)$ is an arbitrary continuous function (note that $\det{\cal B}$ in (\ref{zt}) is evaluated in the space of continuous functions $f^\mu(t)$).
The second term in (\ref{Bquad}) is non-negative, since $-\del_n\del_n'D(\te,\te')$ defines a non-negative quadratic form. Furthermore, if $t(\te)$ develops a discontinuity, $\int\limits_0^{t(\te)} \varphi(t)dt$ acquires a discontinuity as well. Also, $\del_n\del_n'D(\te,\te')$
is a highly singular distribution (due to the singularities of the Dirichlet Green function) \cite{fradkin}. In particular, $\del_n\del_n'D(\te,\te')$ produces infinity if it is contracted with a function that is not continuous. Therefore, for discontinuous $t(\te)$, the second term in (\ref{Bquad}) will become infinitely large and positive, and it will be possible to choose $z(t)$, such that
$\int z(t)dt$ is arbitrarily large and negative, without introducing negative
eigenvalues to the quadratic form (\ref{Bquad}). In other words, it will be possible to deform the integration contour of $z(t)$ and make $\exp[\int z(t) dt]$ arbitrarily close to 0, without ever crossing the singularities of
$\det{\cal B}$ (all the eigenvalues of $\cal B$ are kept non-negative).
} 
arbitrarily far to the left in the complex $z$ plane. Then the integrand will vanish due to the last factor in the second line of (\ref{zt}). This is as it should be, since discontinuous worldsheets do not give any contribution to the original path integral (\ref{master}).

\subsection{Recoil perturbation theory}

Let us now examine how the worldline description of D0-branes works in the recoil regime, i.e., for closed strings scattering off a D0-brane. Since the mass of the D0-brane diverges in the limit $g_{st}\to 0$, it can be treated as static to the lowest order in $g_{st}$ (if we keep the momenta of the incident closed strings fixed as $g_{st}\to0$), and the corrections due to the motion of the D0-brane's center-of-mass (i.e., recoil) will appear as a perturbative expansion in powers of $g_{st}$. This is the familiar recoil perturbation theory. In the present treatment, we shall generate the perturbative expansion by constructing a suitable expansion of the ``effective action'' functional in the path integral (\ref{zt}):
\be
\ber{l}
\dsty S_{eff}[z(t),t(\te)]=-\int z\,dt\vspace{2mm}\\
\dsty\hspace{0.5cm}+\frac14\int \Big(p_1^\mu-\int\limits_0^t d(\tilde t,\s) J^\mu(\s) d\tilde t\ds\Big) {\cal B}^{-1}(t,t')\Big(p_{1\mu}-\int\limits_0^{t'} d(\tilde t',\s') J_\mu(\s')d\tilde t'\ds'\Big)dtdt'.
\eer
\ee
Our intention will be to compute the next-to-leading order term of the disk scattering amplitude and show that this correction contains a divergence of the form needed to cancel the annular divergence (\ref{appendixC}).

There is a difficulty one encounters in handling the effective action $S_{eff}$. It is a most straightforward approach to try to construct a Taylor-like expansion of $S_{eff}$ in powers of $t(\te)$ around $t(\te)=\cnst$:
\be
\begin{array}{l}
\dsty S_{eff}(\cnst+t(\te))\vspace{2mm}\\
\dsty\hspace{1cm}=S_{eff}(\cnst)+\int d\te \left.\frac{\de S_{eff}}{\de t(\te)}\right|_{\mbox{\tiny const}}t(\te)+\frac12\int d\te d\te' \left.\frac{\de^2 S_{eff}}{\de t(\te)\de t(\te')}\right|_{\mbox{\tiny const}}t(\te)t(\te')+\cdots\ .
\end{array}
\ee
This strategy comes to mind in immediate relation to the computational techniques most commonly used in $\s$-models, and it has been employed in \cite{hirano-kazama} for the purposes we are presently pursuing here. Unfortunately, such an expansion does not exist. In \cite{hirano-kazama}, various $\zeta$-function prescriptions have been devised to deal with the infinities arising when one tries to 
implement the Taylor-like expansion, but, as a result, the simple divergent structure of the type needed to cancel (\ref{appendixC}) was lost.

The origin of the above complication can be traced back to the non-analytic properties of the worldlines in the path integral (\ref{0+1}). Indeed, for non-analytic $f^\mu(t)$ (most worldlines are fractal and therefore non-differentiable \cite{ambjoern}) the ``effective action'' in (\ref{0+1}) does not admit a Taylor-like expansion in $t(\te)$ around {\it any} configuration of $t(\te)$.
The integration over $f^\mu(t)$ improves the situation considerably: the resulting effective action can be expanded around any $t(\te)\ne\cnst$,  but the non-analyticity still survives\footnote{One can observe the failure of
the Taylor-like expansion around $t(\te)=\cnst$ by inspecting the second functional derivative $\de^2S_{eff}/\de t(\te_1)\de t(\te_2)$. Applying the Leibniz rule, one will obtain, among others, the following term:
\be
\frac12\int dtdt' {\cal B}^{-1}(t,t')\,\frac{\de}{\de t(\te_1)}\Big[p_1^\mu-\int\limits_0^t d(\tilde t,\s) J^\mu(\s) d\tilde t\ds\Big]\, \frac{\de}{\de t(\te_2)}\Big[p_{1\mu}-\int\limits_0^{t'} d(\tilde t',\s') J_\mu(\s')d\tilde t'\ds'\Big],
\ee
which, after evaluating the functional derivatives, can be rewritten as
\be
\frac12\,{\cal B}^{-1}\left(t(\te_1),t(\te_2)\right) \int \del_nD(\te_1,\s) J^\mu(\s) \ds\, \int \del_nD(\te_2,\s') J_\mu(\s') \ds'.
\ee
If $t(\te)=t_0=\cnst$, the term in $\cal B$ containing ${\cal N}(t,t')$ becomes
negligible, and we have
\be
{\cal B}^{-1}\left(t(\te_1),t(\te_2)\right)=\frac1{z(t(\te_1))}\,\de(t(\te_1)-t(\te_2))=
\frac1{z(t_0)}\,\de(0)=\infty
\ee
Therefore, the particular term we are considering will give an infinite contribution to the second functional derivative of $S_{eff}$, and the
Taylor-like expansion of $S_{eff}$ will not exist.} for worldsheets whose boundary shrinks to a single point.

Luckily, the Taylor-like expansion is not the only way to generate a sensible perturbation theory. Appearing as insertions in a Gaussian path integral, the exponentials of $t(\te)$ are just as tractable as powers of $t(\te)$. We shall therefore resort to a combination of a Taylor-like and a Fourier-like expansion. We first introduce $z(t)=z_0+\de z(t)$ and
\be
A(t,t')=z_0\,\de(t-t'),\qquad B(t,t')=\de z(t)\,\de(t-t')+\frac{1}{4\pi\alpha'}\int\limits_0^t d\tilde t \int\limits_0^{t'} d\tilde t'\,{\cal N}(\tilde t,\tilde t'),
\ee
such that ${\cal B}(t,t')=A(t,t')+B(t,t')$, and expand formally
\begin{equation}
{\cal B}^{-1}=\frac1A-\frac1AB\frac1A+\frac1AB\frac1AB\frac1A+\cdots\ .
\label{AB}
\end{equation}
Note that, at this point, the value of $z_0$ still needs to be specified, and it shall be prudent to choose it in such a way that the term linear in $\de z(t)$ is absent in the resulting expansion of $S_{eff}$. 

The ultimate goal of this expansion is to recast the effective action in the form
\beq
S_{eff}=S_{Gauss}[t(\te),\de z(t)]+S_{pert}[t(\te),e^{i\omega t(\te)},\de z(t)],
\label{pert}
\eeq
where $S_{Gauss}$ is a quadratic Gaussian functional, and $S_{pert}$ can be treated perturbatively (i.e., expanding the exponential of $S_{pert}$ in Taylor series will produce a well-defined power series expansion in string coupling, after the path integral is performed). The above representation highlights the peculiar circumstance that the exponentials of $t(\te)$ will be explicitly retained in $S_{pert}$ alongside with powers of $t(\te)$. When $\exp(-S_{pert})$ is expanded in Taylor series, such exponentials will appear as insertions in a Gaussian path integral, and, of course, integrating them will pose no more difficulty than integrating powers of $t(\te)$. On the other hand, expanding $e^{i\omega t(\te)}$ in powers of $t(\te)$ (to induce a polynomial structure
$S_{pert}$, as would be most commonly done in the various perturbative treatments) would result in a miserably ill-defined perturbative expansion, as it has been already remarked.

We can now implement the expansion (\ref{AB}) in $S_{eff}$ and isolate the following terms contributing to $S_{Gauss}$:
\beq
\begin{array}{l}
\dsty S_{Gauss}^{(1)}=\left(\frac{p^2}{4z_0}-z_0\right)T;\vspace{4mm}\\
\dsty S_{Gauss}^{(2)}=-\frac{p^2}{16\pi\alpha'z_0^2}\int dtdt'\int\limits_0^t d\tilde t \int\limits_0^{t'} d\tilde t'\,{\cal N}(\tilde t,\tilde t')\vspace{2mm}\\
\dsty\hspace{4cm}=\frac{p^2}{16\pi\alpha'z_0^2}\int t(\te)\del_n\del_{n'}D(\te,\te')t(\te') d\te d\te';\vspace{4mm}\\
\dsty S_{Gauss}^{(3)}=\frac{p^2}{4z_0^3}\int{\de z^2}dt;\vspace{4mm}\\
\dsty S_{Gauss}^{(4)}=-\frac{1}{2z_0}p_1^\mu\int dt\int\limits_0^t d(\tilde t,\s) J^\mu(\s) d\tilde t\ds\vspace{2mm}\\
\dsty\hspace{2cm}=-\frac{1}{2z_0}p_1(p_1+p_2)T+\frac{1}{2z_0}p_1^\mu\int t(\te)\del_nD(\te,\s) J_\mu(\s) d\te\ds,
\end{array}
\label{seffgauss}
\eeq
where we have introduced $p^2=p_1^2=p^2_2$ (note that the values of $p_1^2$ and $p_2^2$ can be set equal before applying the reduction formula, without affecting the values of the on-shell amplitudes).

The first term in (\ref{seffgauss}) merely provides for the correct pole structure. It will disappear after integration over $T$ and application of the reduction formula. The remaining three
terms define a Gaussian integral with respect to $\de z(t)$ and $t(\te)$. (In terms of the expansion (\ref{AB}), $S_{Gauss}^{(1)}$ and $S_{Gauss}^{(4)}$ arise from the $A^{-1}$ term, $S_{Gauss}^{(2)}$ arises from the $-A^{-1}BA^{-1}$ term, and $S_{Gauss}^{(3)}$ arises from the $A^{-1}BA^{-1}BA^{-1}$ term.) The following relation has been used for removing the shorthands ${\cal N}(t,t')$ and $d(\te,\s)$ and bringing $S_{Gauss}$ into a manifestly quadratic form:
\be
\int\limits_0^T d\tilde t\int\limits_0^{\tilde t} dt\,\de(t-t(\te))=T-t(\te).
\ee

When writing down the Gaussian part of the effective action (\ref{seffgauss}), we did not include the term proportional to $\de z(t)$. Such term will only be absent if $1+p^2/4z_0^2=0$.
Evidently, this equation has two solutions 
\beq
z_0=\pm i\sqrt{p^2}/2.
\label{z0}
\eeq
Note that treating $\de z(t)$ in perturbation theory effectively amounts to a saddle point evaluation of the worldline path integral. For that reason, we should take a (coherent) sum of the contributions from the two (complex conjugate) saddle points $z_0=\pm i\sqrt{p^2}/2$ when evaluating the amplitude.

At this point, it is instructive to list the following substitution rules,
which allow one to easily identify the powers of $M$ corresponding to the
various terms in the perturbative expansion (\ref{pert}): $p\to M$, $J\to 1$, $z_0\to M$, $\de z\to \sqrt{M}$, $t(\te)\to 1$. The first two rules follow
directly from kinematics (we have chosen the power of $M$ corresponding to
$p_0$, the largest component of $p_\mu$, since it would contribute to a generic Lorentz-invariant expression). The rule for $z_0$ reflects the (on-shell) value of (\ref{z0}), while the last two rules can be deduced by inspecting the
Gaussian action (\ref{seffgauss}).

\subsection{The next-to-leading order corrections}

We shall now proceed with the analysis of the perturbation part $S_{pert}$ of the effective action $S_{eff}$ and evaluation of the next-to-leading order contribution to the disk scattering amplitude. Inspecting further the expansion (\ref{AB}), one identifies the following contributions to $S_{pert}$ relevant to the computation of the next-to-leading order corrections:
\be
\begin{array}{c}
\dsty F_1=\frac{1}{4z_0}\int dt\left(\int\limits_0^{t} d(\tilde t,\s) J^\mu (\s) d\tilde t\ds\right)^2;\vspace{2mm}\\
\dsty F_2=\frac{1}{2z_0^2}\,\frac{1}{4\pi\alpha'}\,\,p_1^\mu\int dtdt'\int\limits_0^{t} d\tilde t \int\limits_0^{t'} d\tilde t'\,{\cal N}(\tilde t,\tilde t')\int\limits_0^{t'}d\tilde t''\ds\,
d(\tilde t'',\s) J_\mu (\s);\vspace{2mm}\\
\dsty F_3=\frac{p^2}{4z_0^3}\left(\frac{1}{4\pi\alpha'}\right)^2\int dt\,dt'\,dt''\int\limits_0^{t} d\tilde t \int\limits_0^{t'} d\tilde t'\,{\cal N}(\tilde t,\tilde t')\int\limits_0^{t'} d\widetilde{\tilde t'}\int\limits_0^{t''} d\tilde t''\,{\cal N}(\widetilde{\tilde t'},\tilde t'');\vspace{2mm}\\
\dsty F_4=\frac{1}{2z_0^2}\,p_1^\mu\int dt\,\de z(t)\int\limits_0^{t} d(\tilde t,\s) J_\mu (\s) d\tilde t\ds;\vspace{2mm}\\
\dsty F_5=\frac{p^2}{2z_0^3}\,\,\frac{1}{4\pi\alpha'}\int dt\,dt'\,\de z(t)\int\limits_0^{t} d\tilde t \int\limits_0^{t'} d\tilde t'\,{\cal N}(\tilde t,\tilde t');\vspace{2mm}\\
\dsty F_6=-\frac{1}{2z_0^3}\,p_1^\mu\int dt\, \de z(t)^2\int\limits_0^{t} d(\tilde t,\s) J_\mu (\s) d\tilde t\ds.
\end{array}
\ee
(In terms of the expansion (\ref{AB}), $F_1$ is contained in $A^{-1}$, $F_2$ and $F_4$ are contained in $-A^{-1}BA^{-1}$, and $F_3$, $F_5$ and $F_6$ are contained in $A^{-1}BA^{-1}BA^{-1}$.)

Once the integration over $\de z(t)$ has been performed with the quadratic Gaussian form $S_{Gauss}^{(3)}$ of (\ref{seffgauss}), the situation simplifies considerably. The contraction of $F_4$ and $F_5$ cancels $F_2$. The contraction of two copies of $F_5$ cancels $F_3$. The integral of $F_6$ over $\de z(t)$ is divergent, but it will merely combine with $S_{Gauss}^{(4)}$ and renormalize the value of $z_0$ adding divergent higher order corrections\footnote{Note that such renormalizations would be necessary even if one tried to treat perturbatively the integration over $\de z(t)$ for a free particle, and they do not have anything to do with the problem of recoil per se.} to (\ref{z0}). Finally, the only non-trivial correction arising at the next-to-leading order will come from $F_1$ and from the contraction of two copies of $F_4$:
\beq
\Delta S_{NLO}[t(\te)]=\frac{1}{4z_0}\left(\de^{\mu\nu}-\frac{p_1^\mu p_1^\nu}{p^2}\right)\int dt\int\limits_0^{t} d(\tilde t,\s) J_\mu (\s) d\tilde t\ds \int\limits_0^{t} d(\tilde t',\s') J_\nu (\s') d\tilde t'\ds'.
\label{SNLO}
\eeq
At this point, it is convenient to eliminate the shorthand $d(t,\s)$ by means of the relation
\beq
\int\limits_0^T dt\int\limits_0^{t} d\tilde t\, \de(\tilde t - t(\te)) \int\limits_0^{t}d\tilde t'\, \de(\tilde t' - t(\te'))=T-\frac{t(\te)+t(\te')}2-G_{kick}\left(t(\te')-t(\te)\right),
\label{intkick}
\eeq
where
\be
G_{kick}(t)=\frac{|t|}{2}
\ee
is the ``causal'' Green function of a free particle (whose formal appearance should indeed be very welcome in a treatment of the recoil problem).

After we substitute (\ref{intkick}) into (\ref{SNLO}), it is convenient to absorb into a re-definition of $S_{Gauss}$ the contributions originating from the terms $T$ and $-(t(\te)+t(\te'))/2$ of (\ref{intkick}). We shall thus introduce
\beq
\ber{l}
\dsty\Delta\tilde S_{NLO}[t(\te)]=-\frac{1}{4z_0}\left(\de^{\mu\nu}-\frac{p_1^\mu p_1^\nu}{p^2}\right)\vspace{3mm}\\
\dsty\hspace{1cm}\times\int d\te\,d\te'\,\ds\,\ds'\,G_{kick}\left(t(\te')-t(\te)\right)\, \del_nD(\te,\s) J_\mu (\s)\,\del_{n'}D(\te',\s')J_\nu (\s')
\eer
\label{SNLOfinal}
\eeq
and%
\footnote{Note that $p_2$ is {\it minus} the momentum of the outgoing (final state) D0-brane in our conventions. Therefore, $p_1-p_2$ in (\ref{tildeGauss}) is the average momentum of the initial and final state of the D0-brane.
}
\beq
\tilde S_{Gauss}^{(4)}=\frac{1}{2z_0}\left(\frac{p_1-p_2}2\right)^\mu\int t(\te)\del_nD(\te,\s) J_\mu(\s) d\te\ds,
\label{tildeGauss}
\eeq
satisfying the relation $\tilde S_{NLO}+\tilde S_{Gauss}^{(4)}=S_{NLO}+ S_{Gauss}^{(4)}$ (up to terms of order $g_{st}^2$).

Turning back to the original expressions for the scattering amplitude (\ref{reduct1}) and (\ref{zt}), at next-to-leading order of recoil perturbation theory, we have to compute%
\footnote{We shall not worry about the overall coefficient here. It is (in any case) not fixed by the path integral itself and needs to be determined from unitarity considerations. Note however that the overall coefficient in (\ref{reductpert}) does not influence the ratio of the leading and the next-to-leading order contributions to the disk amplitude, which is the primary goal of our computations.
} 
the expression:
\beq
\ber{l}
\dsty\left<p_2|p_1\right>_J^{(1)}\sim\de(p_1+p_2+\sum k_n)\lim\limits_{p^2\to -M^2}
\left(p^2+M^2\right)^2\exp\left[-\pi\alpha'\int J^\mu(\s)D(\s,\s')J_\mu(\s')\ds\ds'\right]\vspace{2mm}\\
\dsty\hspace{2cm}\times
\int dT \exp\left[-MT-S_{Gauss}^{(1)}\right] \int \D t(\te)\left(-\Delta \tilde S_{NLO}\right)\exp\left[-S_{Gauss}^{(2)}-\tilde S_{Gauss}^{(4)}\right].
\eer
\label{reductpert}
\eeq

It is important to understand first how the reduction formula works (even though this aspect of the computation is purely kinematic and by no means specific to recoil). An essential circumstance to realize is that the integrand in (\ref{reductpert}) does not depend on the constant mode of $t(\te)$ ($\bar t= \int t(\te)d\te$). Because of that, the integration over the constant mode of $t(\te)$ will produce a factor of $T$, up to corrections from the fact that the integration domain is determined by the full profile of the worldsheet boundary, not only the constant mode. Since the typical size of the worldsheet boundary is $\sqrt{\alpha'}$ (and does not depend on $T$), these corrections grow more slowly than $T$. In other words,
\be
\ber{l}
\dsty\left<p_2|p_1\right>_J^{(1)}\sim\de(p_1+p_2+\sum k_n)\lim\limits_{p^2\to -M^2} \left(p^2+M^2\right)^2\vspace{2mm}\\
\dsty\hspace{0.5cm}\times \exp\left[-\pi\alpha'\int J^\mu(\s)D(\s,\s')J_\mu(\s')\ds\ds'\right]\int dT\left(T+o(T)\right)\exp\left[-MT-S_{Gauss}^{(1)}\right]\vspace{2mm}\\
\dsty\hspace{4cm} \times\int \D \tilde t(\te)\left(-\Delta \tilde S_{NLO}\right)\exp\left[-S_{Gauss}^{(2)}-\tilde S_{Gauss}^{(4)}\right],
\eer
\ee
where $\tilde t(\te)$ is orthogonal to the constant mode ($\int d\te\,\tilde t(\te)=0$), and the contribution $o(T)$ comes from the endpoints of the worldline. If we now keep in mind that, with (\ref{z0}), $S_{Gauss}^{(1)}$ of (\ref{seffgauss}) becomes
\be
S_{Gauss}^{(1)}=\mp i\sqrt{p^2}T,
\ee
it is only left to notice that the integral
\beq
\int dT\,T\,\exp\left[-MT-S_{Gauss}^{(1)}\right]=\int dT\,T\,\exp\left[-(M\mp i\sqrt{p^2})T\right]=\frac1{(M\mp i\sqrt{p^2})^2}
\label{doublepole}
\eeq
produces a double pole necessary for the application of the reduction formula, whereas the integral
\be
\int dT\,o(T)\,\exp\left[-MT-S_{Gauss}^{(1)}\right]=\int dT\,o(T)\,\exp\left[-(M\mp i\sqrt{p^2})T\right]
\ee
will produce a weaker singularity, which will not survive the amputation of external legs. We therefore arrive at the following expression for the next-to-leading order correction:
\beq
\ber{l}
\dsty\left<p_2|p_1\right>_J^{(1)}\sim\de(p_1+p_2+\sum k_n)\,\exp\left[-\pi\alpha'\int J^\mu(\s)D(\s,\s')J_\mu(\s')\ds\ds'\right]\vspace{3mm}\\
\dsty\hspace{2cm}\times\int \D \tilde t(\te)\left(-\Delta \tilde S_{NLO}\right)\exp\left[-S_{Gauss}^{(2)}-\tilde S_{Gauss}^{(4)}\right].
\eer
\label{reduced}
\eeq
Note that, of the two saddle points of $z_0$ given in (\ref{z0}), only one will produce the appropriate singularity in (\ref{doublepole}), whereas the other one will leave the denominator non-vanishing, and hence will give a zero contribution after the application of the reduction formula. The on-shell value of $z_0$ corresponding to the saddle point that gives a non-trivial contribution to the scattering amplitude is given by
\be
z_0=\frac{M}2
\ee
and this is the value that should be used in (\ref{reduced}).

If we now specialize to the Lorentz frame in which the D0-brane moves with (spatial) momentum $\vec{P}/2$ before recoil, and (spatial) momentum $-\vec{P}/2$ after recoil,
\be
p_1=\left(iM+\frac{i}{2M}\hspace{-1mm}\left(\frac{\vec{P}}{2}\right)\hspace{-2mm}\rule{0mm}{6.5mm}^2\hspace{1mm},\hspace{1mm}\frac{\vec{P}}{2}\hspace{1mm}\right),\hspace{2cm}p_2=-\left(iM+\frac{i}{2M}\hspace{-1mm}\left(\frac{\vec{P}}{2}\right)\hspace{-2mm}\rule{0mm}{6.5mm}^2\hspace{1mm},\hspace{1mm}-\frac{\vec{P}}{2}\hspace{1mm}\right),
\ee
the expression (\ref{reduced}) can be rewritten (up to terms of order $g_{st}^2$) in the form
\beq
\ber{l}
\dsty\left<p_2|p_1\right>_J^{(1)}\sim\de\left(\sum k^0_n\right)\,\de\left(\vec{P}+\sum \vec{k}_n\right)\exp\left[-\pi\alpha'\int J^\mu(\s)D(\s,\s')J_\mu(\s')\ds\ds'\right]\vspace{3mm}\\
\dsty\hspace{5mm}\times\int \D \tilde t(\te)\,\frac{1}{2M}\int d\te\,d\te'\,\ds\,\ds'\,G_{kick}\left( \tlt(\te')-\tlt(\te)\right)\, \del_nD(\te,\s) J^i (\s)\,\del_{n'}D(\te',\s')J^i (\s')\vspace{3mm}\\
\dsty\hspace{1cm}\times\exp\left[\frac{1}{4\pi\alpha'}\int \tlt(\te)\del_n\del_{n'}D(\te,\te')\tlt(\te') d\te d\te'- i\int \tlt(\te)\del_nD(\te,\s) J^0(\s) d\te\ds\right]
\eer
\label{correction}
\eeq
with the index $i$ running over the Dirichlet directions only and the path integration performed exclusively over $\tlt(\te)$ orthogonal to the constant mode.%
\footnote{One should keep in mind that, if $D(\s,\s')$ is defined by $\Delta D(\s,\s')=-\de(\s-\s')$, the quadratic form $\int \xi(\te)\del_n\del_{n'}D(\te,\te')\xi(\te') d\te d\te'$ is negative, with the constant mode being the only zero mode. This can be easily seen by considering
\be
-\int \xi(\te)\del_n\del_{n'}D(\te,\te')\xi(\te') d\te d\te'=\int X_\xi(\te)\del_n X_\xi(\te)d\te=\int \left(\nabla X_\xi\right)^2d\s\ge 0
\ee
with $X_\xi$ being the solution of the Laplace equation satisfying $X_\xi(\te)=\xi(\te)$.
}

If we further Fourier-transform $G_{kick}$ according to
\be
G_{kick}(t)=-\frac{1}{4\pi}\int\left(\frac{1}{(\omega+i0)^2}+\frac{1}{(\omega-i0)^2}\right)e^{-i\omega t}d\omega,
\ee
the evaluation of the next-to-leading order correction (\ref{correction}) to the disk amplitude becomes a matter of straightforward Gaussian integration:
\beq
\ber{l}
\dsty\left<p_2|p_1\right>_J^{(1)}\sim\de\left(\sum k^0_n\right)\,\de\left(\vec{P}+\sum \vec{k}_n\right)\exp\left[-\pi\alpha'\int J^\mu(\s)D(\s,\s')J_\mu(\s')\ds\ds'\right]\vspace{3mm}\\
\dsty\hspace{5mm}\times\int \D \tilde t(\te)\,\frac{-1}{8\pi M}\int d\omega\,d\te\,d\te'\,\ds\,\ds'\vspace{3mm}\\
\dsty\hspace{1cm}\times \left(\frac{1}{(\omega+i0)^2}+\frac{1}{(\omega-i0)^2}\right)e^{-i\omega \left( \tlt(\te')-\tlt(\te)\right)}\del_nD(\te,\s) J^i (\s)\,\del_{n'}D(\te',\s')J^i (\s')\vspace{3mm}\\
\dsty\hspace{1cm}\times\exp\left[\frac{1}{4\pi\alpha'}\int \tlt(\te)\del_n\del_{n'}D(\te,\te')\tlt(\te') d\te d\te'- i\int \tlt(\te)\del_nD(\te,\s) J^0(\s) d\te\ds\right].
\eer
\label{correctionFourier}
\eeq

\subsection{Divergence cancellation}

The expression (\ref{correctionFourier}) turns out to be divergent upon closer inspection. This is all for the better, since, if it were finite, there would not be anything within the formalism to cancel the annulus divergence (\ref{appendixC}). It is certainly true that, once the divergences are cancelled, it is the remaining finite part of the amplitude that reflects the physical contents of the theory. Yet, since it is the divergence cancellation that is of principal importance for us here, we shall not be paying much attention to the finite (physical) part of (\ref{correctionFourier}), but shall concentrate on the divergence instead.

Before we proceed with the divergence computation, let us first note that the leading order disk amplitude is given in the worldline formalism by
\beq
\ber{l}
\dsty\left<p_2|p_1\right>_J^{(0)}\sim\de(p_1+p_2+\sum k_n)\lim\limits_{p^2\to -M^2} \left(p^2+M^2\right)^2\exp\left[-\pi\alpha'\int J^\mu(\s)D(\s,\s')J_\mu(\s')\ds\ds'\right]\vspace{2mm}\\
\dsty\hspace{2cm}\times \int dT \exp\left[-MT-S_{Gauss}^{(1)}\right] \int \D t(\te)\exp\left[-S_{Gauss}^{(2)}-\tilde S_{Gauss}^{(4)}\right]\vspace{6mm}\\
\dsty\hspace{5mm}\sim\de\left(\sum k^0_n\right)\,\de\left(\vec{P}+\sum \vec{k}_n\right)\exp\left[-\pi\alpha'\int J^\mu(\s)D(\s,\s')J_\mu(\s')\ds\ds'\right]\vspace{3mm}\\
\dsty\hspace{5mm}\times\int \D \tilde t(\te)\exp\left[\frac{1}{4\pi\alpha'}\int \tlt(\te)\del_n\del_{n'}D(\te,\te')\tlt(\te') d\te d\te'- i\int \tlt(\te)\del_nD(\te,\s) J^0(\s) d\te\ds\right].
\eer
\label{(0)}
\eeq
These expressions%
\footnote{Note that, should one be willing to recover the disk amplitude in its most conventional form from this representation, one should keep in mind the relations between the Neumann and Dirichlet Green functions (see the appendices of \cite{eoe}). The result of the integration over $\tlt(\te)$ in (\ref{(0)}) is of precisely such form that it will correct $J^0(\s)D(\s,\s')J_0(\s')$ to $J^0(\s)N(\s,\s')J_0(\s')$. The formal presence of the Neumann Green function will thus be restored for the Neumann directions, as was to be expected.
} 
are direct analogs of (\ref{reductpert}) and (\ref{correction}), without the insertion of the perturbative corrections coming from expanding $\exp[-S_{pert}]$.

With the above formula for the leading order disk amplitude, we can represent the expression (\ref{correctionFourier}) for the next-to-leading order correction in the following convenient form:
\beq
\ber{l}
\dsty\left<p_2|p_1\right>_J^{(1)}=\left<p_2|p_1\right>_J^{(0)}\frac{-1}{8\pi M}\int d\omega\,d\te\,d\te'\,\ds\,\ds'\left(\frac{1}{(\omega+i0)^2}+\frac{1}{(\omega-i0)^2}\right)\vspace{3mm}\\
\dsty\hspace{5mm}\del_nD(\te,\s) J^i (\s)\,\del_{n'}D(\te',\s')J^i (\s')\exp\left[-\pi\alpha'\left(\omega^2\left(N(\te,\te)+N(\te',\te')-2N(\te,\te')\right)\phantom{\int}\right.\right.\vspace{2mm}\\
\dsty \left.\left.\hspace{3cm}+2\omega\int \left(N(\te',\tilde\te)-N(\te,\tilde\te)\right) \del_nD(\tilde\te,\s)J^0(\s) d\tilde\te\ds\right)\right],
\eer
\label{(1)}
\eeq
where $N$ denotes the Neumann Green function and we have used\footnote{In transition from (\ref{correctionFourier}) to (\ref{(1)}), we performed the Gaussian integration in (\ref{correctionFourier}), and then used the relation
(\ref{DirichletNeumann}) to replace the kernel of the inverse of the quadratic Gaussian form in (\ref{correctionFourier}) with the Neumann Green function. It is very natural that the Neumann Green function re-appeared at the end of our derivations, since we have been considering a D0-brane, and thus had one Neumann direction in the problem from the very beginning.} the relation \cite{eoe}
\be
\int\deln\del_{\tilde n} D(\te,\tte)N(\tilde\te,\te')\,d\tte=
-\tilde\de(\te-\te')
\label{DirichletNeumann}
\ee
(where $\tilde\de(\te)$ is the zero-mode-orthogonal $\de$-function).

If we now substitute into (\ref{(1)}) the regularized value of the singularity of the Neumann function\footnote{In our conventions, $\Delta N(\s,\s')=-\tilde\de(\s-\s')$, where $\tilde\de(\s-\s')$ is the zero-mode-orthogonal $\de$-function. Then, the singularities of the Neumann Green function (on an arbitrary worldsheet) can be extracted from the expression for the Neumann Green function on a flat semi-infinite worldsheet:
\be
N(z,z')=-\frac1{4\pi}\log|z-z'|^2-\frac1{4\pi}\log|z-\bar z'|^2
\ee
(where we have used the complex worldsheet parametrization).}
\be
N(\te,\te)\to N(\te,\te+\de)=-\frac{1}{\pi}\log\de
\ee
(with $\de$ being the worldsheet ``minimal distance'' cut-off), the relevant $\omega$-integral becomes:
\beq
\ber{l}
\dsty\int d\omega\left(\frac{1}{(\omega+i0)^2}+\frac{1}{(\omega-i0)^2}\right)\exp\left[-\pi\alpha'\left(\omega^2\left(N(\te,\te)+N(\te',\te')-2N(\te,\te')\right)\phantom{\int}\right.\right.\vspace{2mm}\\
\dsty \left.\left.\hspace{3cm}+2\omega\int \left(N(\te',\tilde\te)-N(\te,\tilde\te)\right) \del_nD(\tilde\te,\s)J^0(\s) d\tilde\te\ds\right)\right]\vspace{4mm}\\
\dsty=\int d\omega\left(\frac{1}{(\omega+i0)^2}+\frac{1}{(\omega-i0)^2}\right)e^{2\alpha'\omega^2\log\de}\vspace{2mm}\\
\dsty \hspace{1cm}\exp\left[2\pi\alpha'\left(\omega^2N(\te,\te')-\omega\int \left(N(\te',\tilde\te)-N(\te,\tilde\te)\right) \del_nD(\tilde\te,\s)J^0(\s) d\tilde\te\ds\right)\right]\vspace{6mm}\\
\dsty\hspace{4cm}=-4\sqrt{2\pi\alpha'|\log\de|}+\mbox{finite terms}.
\eer
\label{kappa}
\eeq
Note that the terms of the fourth line of (\ref{kappa}) do not contribute into the divergent part of the $\omega$-integral, as can be made manifest by the rescaling $\omega'=\omega\sqrt{2\alpha'|\log\de}|$. The remaining integral can be evaluated using
\be
\int\frac{1}{(\omega\pm i0)^2}\,e^{-a\omega^2}d\omega=-2\sqrt{\pi a}+\mbox{const}.
\ee

Finally, substituting (\ref{kappa}) into (\ref{(1)}) and keeping in mind that
\be
\int \del_nD(\te,\s) J (\s)d\te d\s=p_1+p_2=P,
\ee
we conclude that the divergent part of the next-to-leading order correction to the disk amplitude has the form
\beq
\left<p_2|p_1\right>_J^{(1;div)}=\sqrt{|\log\de|}\,\,\frac{P^2}{M}\,\,\sqrt{\frac{\alpha'}{2\pi}}\,\,\left<p_2|p_1\right>_J^{(0)}.
\label{diskdiv}
\eeq

We should now compare (\ref{diskdiv}) with the annulus divergence (\ref{appendixC}). The derivation of (\ref{appendixC}) has been given within the worldsheet CFT rather than the worldline formalism. However, it should be clear on physical grounds, and can be deduced by inspecting (\ref{(0)}), that at lowest order in $g_{st}$ on each given worldsheet, the worldline formalism will merely append the Dirichlet momentum conservation $\de$-function to the result coming from the worldsheet CFT computation. For that reason, (\ref{appendixC}) implies that, in the worldline formalism, to order $g_{st}$, the divergent part of the annulus amplitude is given by
\beq
\left<p_2|p_1\right>_J^{(annulus;div)}=-\sqrt{|\log\eps|}\,\,\frac{P^2}{2M}\,\,\sqrt\frac{\alpha'}{\pi}\,\,\left<p_2|p_1\right>_J^{(0)}.
\label{annulusdiv}
\eeq

In order to compare (\ref{diskdiv}) and (\ref{annulusdiv}), it is important to establish a relation between the minimal worldsheet distance cut-off $\de$ and
the minimal gluing parameter cut-off $\eps$. To do so, it suffices to remember that the minimal cross-section of the strip generated by the plumbing fixture construction \cite{polchinski-fischler-susskind} with gluing parameter $\eps$ is proportional to $\sqrt{\eps}$. 
It is therefore natural to identify $\eps\sim\de^2$.
With such an identification,\footnote{
A possible factor $a$ in $\eps=a\de^2$ does not affect (\ref{diskdiv})
and (\ref{annulusdiv}). Furthermore, the $a$-dependence in the finite part of the annulus amplitude is suppressed by $1/\sqrt{|\log\eps|}$, so that the physical amplitude is unambiguously determined in the $\eps\to0$ limit. We thank N.~Nekrasov for pointing this out.}
the divergences (\ref{diskdiv}) and (\ref{annulusdiv}) indeed cancel each other.
We have thus shown (technically, to the order $g_{st}$) that introducing dynamical D0-brane worldlines automatically cures the infrared divergences
present in the conventional worldsheet description of D0-branes and associated
with recoil.

\section{The bilocal recoil operator}

The worldline formalism provides a very natural way to deal with the translational mode of the D0-brane. Nevertheless, in previous work, deformations of the D0-brane CFT with recoil operators have been given a greater amount of attention. It would therefore be beneficial to complete the picture and show that our worldline results can be mimicked by including an appropriate recoil operator in the worldsheet CFT.

To this end, we shall inspect the following bilocal operator:
\beq
V_{bilocal}=\frac1{2M}\int d\te d\te'\,\,G_{kick}\left(X^0(\te)-X^0(\te')\right)
\,:\del_nX^i(\te)\del_{n'}X^i(\te'):\ .
\label{bilocal}
\eeq
Here, once again, $G_{kick}$ is the ``causal'' Green function of the free particle:
\be
G_{kick}(t)=\frac{|t|}2.
\ee
Note also that the normal ordering sign enclosing the operators $\del_nX^i(\te)$ and $\del_{n'}X^i(\te')$ excludes contracting those two operators. On the other hand, $G_{kick}\left(X^0(\te)-X^0(\te')\right)$ is not normal ordered at all. (These subtleties in the definition of $V_{bilocal}$ are justified by the comparison with the worldline formalism we are about to make.)

If one inserts the operator (\ref{bilocal}) into the CFT path integral for a closed string scattering amplitude, performs the integration over $X^i(\s)$, and then, using (\ref{gaussian}), performs the integration over the values of $X^0(\s)$ in the interior of the worldsheet (not the boundary), and further identifies $X^0(\te)\to t(\te)$, one precisely recovers\footnote{Note that the integration over $X^i(\s)$ with Dirichlet boundary conditions turns $\del_nX^i(\te)$ into $\int d\s\,\del_nD(\te,\s) J^i (\s)$, as can be seen by contracting $\del_n X^i$ and $\exp(iJ\cdot X)$ in the Dirichlet path integral (given, for example, by \eq{gaussian} with $\xi(\theta)=0$). The exponential part of (\ref{correction}) is easily recognizable as the same as the exponential part of (\ref{gaussian}) under the substitution $X^0(\te)\to t(\te)$.} the next-to-leading order correction (\ref{correction}) coming from the worldline formalism (except for the spatial momentum conservation $\de$-function, see the discussion below). One therefore concludes that introducing the operator (\ref{bilocal}) as a background correction in the D0-brane CFT precisely mimics (at the next-to-leading order) the contribution of the curved worldlines arising in the dynamical worldline description of recoil. It then follows from the computations in section~4.5 that the operator (\ref{bilocal}) cancels the annular divergence (\ref{appendixC}).

Heuristically, the operator (\ref{bilocal}) can be understood as follows. $\del_{n'}X^i(\te')$ is the operator of momentum density flowing out of the point $\te'$ of the boundary of the worldsheet (integrated over $\te'$, it gives the total transferred Dirichlet momentum $P$). Then,
\beq
\int d\te'\,G_{kick}\left(X^0(\te)-X^0(\te')\right)\del_{n'}X^i(\te')
\label{displace}
\eeq
evaluates the displacement (in the Dirichlet directions) of the point on the D0-brane worldline with the co-ordinate $X^0(\te)$ due to the momentum influx from the closed strings. Finally, the remaining $\del_nX^i(\te)$ actually displaces the point on the D0-brane worldline with the co-ordinate $X^0(\te)$ by the amount (\ref{displace}). Overall, the operator (\ref{bilocal}) deforms the D0-brane worldline precisely in the way that intuitively corresponds to its response to the impact by the closed strings.

Of course, the above description merely serves to relate the background modification we are proposing to the intuitive notion of recoil, and its precise justification comes from the worldline derivations and the corresponding cancellation of divergences.

Note that the operator (\ref{bilocal}) is algebraically quite similar to the recoil operator suggested in \cite{tafjord}, even though they are by no means identical\footnote{The authors of \cite{fischler} attempt to employ a bilocal recoil operator to cancel the annular divergence, but it is different from the correct recoil operator (\ref{bilocal}).}. The operator (\ref{bilocal}) is manifestly time-translation invariant, and it certainly does not imply that the D0-brane is moving (in the act of recoil) along a fixed classical trajectory.
The bilocal structure of the operator (\ref{bilocal}) is also more natural than the local operator in \cite{tafjord} since the annular divergence involves two points on the boundary of the disk (connected by a thin strip).

We end this section with a discussion of a minor (yet instructive) subtlety closely related to the considerations of appendix B. One could na\"\i vely guess that, after the recoil operator is introduced as a deformation of the worldsheet CFT, the resulting finite closed string scattering amplitudes should be referring to a D0-brane moving with a certain velocity in the initial state, and with a different velocity (developed in the course of recoil) in the final state. In other words, one could think that the initial and final states of the D0-brane are momentum eigenstates. This na\"\i ve guess cannot be correct, however, since the amplitude to transition between two momentum eigenstates (in a way that conserves momentum) must be infinite on account of the momentum conservation $\de$-function, whereas the amplitude computed by the worldsheet CFT deformed with the recoil operator is finite by construction. (Note that the presence of the momentum conservation
$\de$-function is closely related to the non-normalizability of the momentum eigenstates.)

The resolution here is that (in close parallel to the considerations of appendix B) the CFT deformed with the recoil operator really describes the transition between normalized wavepacket states (in the limit of spatial extent of the wavepacket going to infinity, and the extent in momentum space shrinking to 0), rather than the (non-normalizable) momentum eigenstates. The difference between these amplitudes is precisely that the limit of the wavepacket amplitude does not contain a momentum conservation $\de$-function (yet it does contain an energy conservation $\de$-function), and this is why the infinitely broad wavepacket states provide the correct way to think about the finite scattering amplitudes, which remain after the CFT divergences have been cancelled with the recoil operator.

\section{Conclusions}

Our aim in this paper has been to reconsider the issue of D0-brane recoil
in bosonic string theory. The various approaches to this problem that have
been previously described in the literature are disparate and lead to incompatible results.

Our primary tool for systematic investigation of the D0-brane recoil has been
the worldline formalism based on introducing dynamical worldlines for D0-branes. In this formalism, the problem of recoil becomes conceptually very simple, since the degree of freedom corresponding to the translational motion of the D0-brane appears in a very explicit fashion.

Using the computational techniques we have developed within the worldline formalism, we have analyzed the recoil scattering amplitudes to next-to-leading order in the string coupling. 
The conventional D0-brane CFT computation of recoil amplitudes is invalidated by an infrared divergent contribution to the annulus diagram.
We have shown that, in the worldline formalism, an (infrared divergent) correction to the string diagram of disk topology automatically cancels the annular divergence.

The cancellation between string diagrams of different topology naturally
suggests a link to the Fischler-Susskind mechanism of divergence cancellation
in string theory. Indeed, we have succeeded to show that, within the conventional
worldsheet CFT, the corrections arising in the worldline formalism can be
imitated by deforming the background of a static D0-brane with a specific
recoil operator. The form of the recoil operator we have found (as an elaboration
of our worldline formalism results) is bilocal and different from the ones previously suggested in the literature.

It would be interesting to show that the bilocal recoil operator deduced
here through an appeal to the worldline formalism is indeed unique, even
from the worldsheet CFT perspective. This would require a careful consideration
of the divergences arising from the various limits in the moduli space, not
only the annulus divergence. Some preliminary remarks to this end have
been made in appendix A. However, at present, we have refrained from
further considerations of this issue.

We have also given an analysis of the recoil process in the DBI formalism.
The advantage of such an approach is that the resummed form of the DBI
action can be used to obtain information about infrared divergences at
all orders in the string coupling (even though the DBI formalism itself
can only be used for a sufficiently small energy of the incident closed
strings). We have seen that the resummed amplitudes are finite and, in place
of the infrared divergences, display the natural momentum conservation features.
Such a resummation pattern is likely to persist to the level of the full string 
theory, even though, as of now, appropriate techniques are lacking to make it
explicit. 

In the DBI analysis, we have also asked what state is implicitly selected by the standard D0-brane CFT. Assuming the corresponding state of the DBI theory to be a Gaussian wavepacket, we have found that the wavepacket has to be sharply localized in momentum space. It will be interesting to check whether this conclusion is consistent with higher order computations in the D0-brane CFT.

\section*{Acknowledgments}

We would like to thank J.~Ambj\o rn for collaboration on the DBI approach to recoil and for many interesting discussions. We also thank C.~Bachas, M.~Gaberdiel, S.~Gukov, A.~Kapustin, A.~Mikhailov, N.~Nekrasov, B.~Pioline, J.~Schwarz and J.~Troost for useful discussions. The work of B.C.\ was supported in part by Stichting FOM, by the Belgian Federal Science Policy Office through the Interuniversity Attraction Pole P5/27, by the European Commission FP6 RTN programme MRTN-CT-2004-005104 and by the ``FWO-Vlaanderen'' through project G.0428.06. B.C.\ thanks the organizers of the First Cambridge-Mitchell Texas Conference and of the 38th International Symposium Ahrenshoop for hospitality during the final stages of this work. S.N.\ was supported in part by the SRC Program of the KOSEF through the Center for Quantum Space-time (CQUeST) of Sogang University with grant number R11 - 2005 - 021. S.N.\ also thanks the Niels Bohr Institute, where part of the present work was done.
\appendix

\section{Comparison with previously advocated approaches to recoil}

In this appendix, we compare our results with three earlier proposals in the literature \cite{tafjord, fischler, hirano-kazama}.

Of the three, the treatment of \cite{hirano-kazama} bears the closest similarity to the considerations of the present paper. Nevertheless, the presentation given in \cite{hirano-kazama} is not satisfactory, as far as the problem of recoil is concerned. Indeed, as has been emphasized in the main text, cancellations between divergences coming from worldsheets of different topologies are essential to the implementation of recoil in string theory. The authors of \cite{hirano-kazama} do not recover the disk divergence in a form that would make it possible to see how it cancels the divergence in the modular integration of the annulus. The derivations described in the present publication are intended to make up for this shortcoming.

The paper \cite{tafjord} is also important for our present considerations, in that it emphasizes the role of the Fischler-Susskind mechanism for the implementation of recoil in string theory. It has already been mentioned section 3 that the version of the Fischler-Susskind mechanism that naturally arises from the worldline formalism bears a strong resemblance to the considerations of \cite{tafjord}. As we could see, a modification of the approach of \cite{tafjord} reconciled it with the considerations building upon the worldline formalism. However, as we shall now argue, the original proposal of \cite{tafjord} does not provide an adequate treatment of recoil.

In order to cancel the modular integration divergence coming from the annulus, the authors of \cite{tafjord} introduce a background correction of the form
\beq
V_{PT}\sim\int d\te\,\del_n X^i(\te) F^i(X^0(\te)), 
\label{PT3}
\eeq
where $F(t)$ is some function and the index $i$ labels the Dirichlet directions. The following concrete choice is made:
\be
F^i(t)=\frac{p_2^i-p_1^i}M\,t\,\Theta(t)
\ee
(where $p_1$, $p_2$ and $M$ are the initial and final momentum and the mass of the D0-brane respectively, and $\Theta(t)$ is the step function). However, as we shall see below, the divergence cancellation only depends on the asymptotic behavior of $F(t)$ in the infinite past and future. Note the absence of normal ordering in (\ref{PT3}).

The physical interpretation of this background correction (the ``recoil operator'') is that the D0-brane starts moving along the trajectory $F(t)$. In particular, the choice of $F(t)$ made in \cite{tafjord} corresponds to the D0-brane abruptly starting to move at the moment $t=0$. The authors then express a minor dissatisfaction with the abruptness of recoil which is manifest in their implementation. However, the main problem with the operator (\ref{PT3}) is not the abruptness by itself, but rather the fact that it singles out a particular moment of time, namely the moment at which the D0-brane starts moving. While the finite part of the amplitude does depend on which moment of time one chooses, all choices are equivalent as far as divergence cancellation goes (as we shall show momentarily). In a classical scattering process, a reasonable choice would be the moment of impact of the scattering process. However, in a quantum mechanical scattering process of states with well-defined momenta, there is no well-defined moment of impact that could single out any particular recoil operator.

If one starts with $F(t)$ satisfying the asymptotic conditions
\be
F^i(t\to-\infty)\sim 0,\qquad F^i(t\to+\infty)\sim v^it+x_*^i,
\ee
the Fourier transform of $F(t)$ can be written in the form
\beq
F^i(\omega)=\frac{v^i}{(\omega+i0)^2}+\frac{x^i_*}{\omega+i0}+\varphi^i(\omega),
\label{Fourier}
\eeq
where $\varphi(\omega)$ is analytic at $\omega=0$. Correspondingly, the operator (\ref{PT3}) can be rewritten in the form
\beq
V_{PT}\sim\int\,d\te\,d\omega\,F(\omega)\,\del_n X^i(\te) \exp\left[-i\omega X_0(\te)\right].
\label{VPT}
\eeq
Let us now insert this operator into the disk amplitude containing the closed string vertex operators $V_1(\s_1)$,$\cdots$,$V_n(\s_n)$ carrying momenta $k_1$,$\cdots$,$k_n$ with $P=\sum k_n$:
\be
\langle V_{PT} V_1(\s_1)\cdots V_n(\s_n)\rangle_{D_2}.
\ee
Performing first the path integral over $X^0$, we obtain
\beq
\int\,d\te\,F^i(P^0)\,\del_n X^i(\te) \exp\left[(P^0)^2\log\eps+P^0 h(\te,\s_n,k_n)\right]\langle V_1(\s_1)\cdots V_n(\s_n)\rangle_{D_2}^{\int X^0}.
\label{intX0}
\eeq
Here, the $(P^0)^2\log\eps$ term in the exponent comes from the self-contraction in the recoil operator, all the contractions among the various vertex operators are symbolically assembled as $\langle V_1(\s_1)\cdots V_n(\s_n)\rangle_{D_2}^{\int X^0}$, and the $P^0 h(\te,\s_n,k_n)$ term comes from the contractions between the recoil operator and the vertex operators. We shall not need the explicit form of the function $h(\te,\s_n,k_n)$. Let us only note that it is non-singular as long as $\s_n$'s stay away from the boundary of the worldsheet. We shall briefly comment below on the important subject of integration over the positions of the vertex operators. Note also that the integration over $\omega$ present in (\ref{VPT}) has disappeared in (\ref{intX0}) due to the energy conservation $\de$-function.

We shall now examine the $\eps\to 0$ limit of the expression (\ref{intX0}). In this limit, it vanishes unless $P^0=0$ and should therefore be thought of as a distribution (made of the $\de$-function and its derivatives) rather than an ordinary function. Hence, to analyze the $\eps\to 0$ limit, we shall examine the convolution of (\ref{intX0}) with an arbitrary smooth function ${\cal G}(P^0)$ admitting a Taylor expansion
\be
{\cal G}(P^0)={\cal G}(0)+P^0{\cal G'}(0)+\cdots\ .
\ee
It is important to keep in mind the expression (\ref{Fourier}) for the function $F$. Only a few terms of the Taylor expansion will contribute to the final answer, since
\be
\int dP^0 \left(P^0\right)^n\exp\left[(P^0)^2\log\eps\right]
\ee
goes to 0 in the limit $\eps\to 0$ if $n\ge 0$. Some other relevant formulas for the integral expressions we shall have to use are
\be
\int dP^0\, \frac{\exp\left[(P^0)^2\log\eps\right]}{\left(P^0+i0\right)^2}\sim\sqrt{|\log\eps|},\qquad\int dP^0\,\frac{\exp\left[(P^0)^2\log\eps\right]}{P^0+i0}\sim 1.
\ee
If we now contract (\ref{intX0}) with ${\cal G}(P^0)$:
\be
\int\,d\te\,dP^0\,{\cal G}(P^0)\,F^i(P^0)\,\del_n X^i(\te) \exp\left[(P^0)^2\log\eps+P^0 h(\te,\s_n,k_n)\right]\langle V_1(\s_1)\cdots V_n(\s_n)\rangle_{D_2}^{\int X^0}
\ee
and use the above integral formulas, we obtain a few distinct terms:
\begin{enumerate}
\item
 The only term that diverges in the limit $\eps\to 0$ is proportional to
\be
\sqrt{|\log\eps|}\,v^i\,{\cal G}(0) \int\,d\te\,\del_n X^i(\te)\,\langle V_1(\s_1)\cdots V_n(\s_n)\rangle_{D_2}^{\int X^0}.
\ee
After integration over $X^i$, this becomes
\be
\sqrt{|\log\eps|}\,{\cal G}(0)\,v^iP^i\,\langle V_1(\s_1)\cdots V_n(\s_n)\rangle_{D_2}.
\ee
The corresponding term in the expression (\ref{intX0}) itself (rather than its convolution with ${\cal G}(P^0)$) is
\be
\sqrt{|\log\eps|}\,\de(P^0)\,v^iP^i\,\langle V_1(\s_1)\cdots V_n(\s_n)\rangle_{D_2}.
\ee
This is precisely the structure that, in \cite{tafjord}, has been claimed to be responsible for cancelling the annular divergence after the final recoil velocity $v$ is adjusted to the value $P/M$. Note, however, that, by itself, this argument does not even determine the velocity $v$, but only its component along $P$. 
\item
Another contribution comes from the $P^0{\cal G'}(0)$ term in the Taylor expansion for ${\cal G}(P^0)$ and is proportional to
\be
\de'(P^0)\,v^iP^i\,\langle V_1(\s_1)\cdots V_n(\s_n)\rangle_{D_2}.
\ee
This term should correct the energy conservation in such a way that the final energy of the D0-brane (proportional to $v^iP^i$) is taken into account.
\item
The last contribution we shall consider here comes from the second term in (\ref{Fourier}). It is proportional to
\be
\de(P^0)\,x_*^iP^i\,\langle V_1(\s_1)\cdots V_n(\s_n)\rangle_{D_2}.
\ee
The essential feature that needs to be highlighted here is that the above expression depends on $x_*$, the parameter that shifts the D0-brane trajectory. Different values of $x_*$ would produce different contributions to the finite part of the closed string scattering amplitude.
\end{enumerate}
However, there is clearly no preferred physical choice for $x_*$. The D0-brane is a quantum particle that does not have a definite position in space as long as its final velocity $v$ is specified. An attempt to implement the approach of \cite{tafjord} thus leads to an apparent absurdity: an infinite-fold ambiguity in the values of the physical amplitudes governed by the value of the fictitious position of the quantum D0-brane.

Finally, let us briefly comment on the paper \cite{fischler}. In its spirit and general attitude to implementing the Fischler-Susskind mechanism, it is similar to the approach of \cite{tafjord}. However, the authors of \cite{fischler} claim that it is necessary to introduce simultaneously two different recoil operators, one of which is bilocal (and resembles, but is not the same as the one we introduced in relation to the worldline formalism), while the other one is a shift of the classical D0-brane trajectory analogous to \cite{tafjord}. From our analysis, we do not see a necessity for two different types of recoil operators. Also, a number of mathematical details appear to be different from what we find. For example, it is claimed in \cite{fischler} that the annular divergence is proportional to $\log\eps/T$ (where $T$ is designated as a ``large time cut-off''), while according to \cite{tafjord} and the present paper, the correct form of the annular divergence is $\sqrt{|\log\eps|}$. 

\section{The DBI picture of D0-brane recoil}

The DBI description is an extremely powerful framework to study infrared
issues for D0-branes, since the resummed non-polynomial form of the DBI action
permits to draw some conclusions valid to all orders in the string coupling.
The picture of recoil one arrives at through such considerations (which is
only valid at low energies, but includes perturbative resummations) is very
complementary to our treatment in the main text, where complete string amplitudes were considered at next-to-leading order in the string coupling.

There is a subtlety one needs to resolve to make an appropriate use of the DBI action for D0-branes. In the DBI action, there is an explicit degree of freedom corresponding to the position of the D0-brane. To compute scattering amplitudes, one needs to specify the initial and final states for this degree of freedom. However, it is not obvious which choice of the initial and final states
corresponds, say, to the standard computations within the (infrared divergent)
D0-brane CFT.

Technically, the question we must ask is what initial and final states for the translational mode of the D0-brane one needs to specify (in the DBI language) in order to make the resulting DBI amplitude mimic the low-energy limit of the worldsheet CFT amplitude without open strings. The answer will not be completely trivial; for example, the na\"\i ve guesses that the state of the D0-brane in the worldsheet CFT is the momentum eigenstate with zero momentum, or a coordinate eigenstate with the D0-brane located at the origin (or a wavepacket localized at the origin with a width of, say, $\sqrt{\alpha'}$) are all incorrect. 

We shall start with the DBI action:
\be
S_{DBI}=-\tau\int d^{p+1}\xi\,e^{-\Phi(X(\xi))}\left\{\det\left[\frac{\del X^\mu}{\del\xi^a}\frac{\del X^\nu}{\del\xi^b}\left(G_{\mu\nu}(X(\xi))+B_{\mu\nu}(X(\xi))\right)+2\pi\alpha'F_{ab}\right]\right\}^{1/2}
\ee
and restrict ourselves, for the sake of convenience, to scattering of one dilaton off a (non-relativistic) D0-brane\footnote{Note that even though there are higher derivative corrections to the DBI action, they would not have a major bearing upon any infrared issues, such as recoil. Indeed, adding derivative interactions to any given process will soften the infrared behavior, and hence will only be able to introduce subleading contributions (compared to the ones coming from the DBI action).}. In this case, the relevant part of the above action is
\beq
M\int dt \left(\frac12 \left(\dot X\right)^2 - \Phi\left(t,X^i(t)\right) + \frac12\left(\Phi(t,X^i(t))\right)^2\right). 
\label{DBI}
\eeq
The interactions of the dilaton and the translational mode involving time derivatives are omitted, as they do not contribute to infrared divergences. 

Let us consider the contribution to the dilaton scattering from the last term in (\ref{DBI}). In the operator language (we work in the interaction picture), it is just
\beq
\left<f,k_2\right|\frac{M}2\int dt \left(\Phi(t,X^i(t))\right)^2\left|i,k_1\right>,
\label{amplitude}
\eeq
where $\left<f\right|$ and $\left|i\right>$ describe the initial and final state of the D0-brane, and $\left<k_2\right|$ and $\left|k_1\right>$ describe the outgoing dilaton of momentum $k_2$ and incoming dilaton of momentum $k_1$. Using
\be
\left<k_2\right|\left(\Phi(t,x^i)\right)^2\left|k_1\right>\sim\exp\left[i(k_2^0-k_1^0)t\right] \exp\left[-i(k_2^i-k_1^i)x^i\right]
\ee
(the overall coefficient of the amplitude will not interest us here), (\ref{amplitude}) can be rewritten as
\beq
\left<f\right|\int dt \exp\left[-i(k_2^i-k_1^i)X^i(t)\right]\exp\left[i(k_2^0-k_1^0)t\right]\left|i\right>.
\label{matrix}
\eeq

We shall simply choose the initial and final states of the D0-brane to be spatially broad Gaussian wave packets (of width $d$ at $t=0$) with momentum centered around $P_1$ and $P_2$, respectively. We shall see that the inverse width of the wave packets should be identified in a particular way with the infrared cut-off imposed in the worldsheet theory.

It is most convenient to compute the matrix element in (\ref{matrix}) in the ``Schr\"odinger'' picture\footnote{The quotation marks highlight that it is the Schr\"odinger picture of the free particle, not of the particle described by (\ref{DBI}).} and in the momentum representation, since the expressions for the Schr\"odinger (momentum) wavefunctions of the initial and final states we have chosen are well-known:
\beq
\ber{c}
\dsty\Psi_f(p,t)\sim d^{D/2}\exp\left[-\frac{(p-P_2)^2d^2}{2}\right]\exp\left[\frac{ip^2t}{2M}\right];\vspace{3mm}\\
\dsty\Psi_i(p,t)\sim d^{D/2}\exp\left[-\frac{(p-P_1)^2d^2}{2}\right]\exp\left[\frac{ip^2t}{2M}\right],
\eer
\label{wave}
\eeq
where $D$ is the number of spatial dimensions. Moreover, in the momentum representation, the operator $\exp[i(k_1-k_2)X]$ simply shifts the wavefunction by $k_1-k_2$. Hence, (\ref{matrix}) becomes
\be
\ber{l}
\dsty\left<f\right|\int dt \exp\left[-i(k_2^i-k_1^i)X^i(t)\right]\exp\left[i(k_2^0-k_1^0)t\right]\left|i\right>\vspace{3mm}\\
\dsty\hspace{2cm}=\int dt \exp\left[i(k_2^0-k_1^0)t\right]\int dp\,\Psi^*_f(p,t)\Psi_i(p+k_2-k_1,t)\vspace{5mm}\\
\dsty\hspace{.5cm}\sim \exp\left[-\frac{(P_2+k_2-P_1-k_1)^2d^2}{2}\right]\vspace{3mm}\\
\dsty\hspace{1.5cm}\times\int dt \exp\left[i\left(k_2^0-k_1^0+\frac{P_2^2}{2M}-\frac{P_1^2}{2M}\right)t\right]\exp\left[-t^2\,\zeta\left(P_i,k_n\right)/d^2\right],
\eer
\ee
where $\zeta$ is some function of the momenta whose precise form will drop out of the final result. The important point is that, as $d$ goes to infinity, the integral over $t$ becomes just the energy conserving $\delta$-function, and the DBI expression for the particular amplitude we are considering (scattering of one dilaton off the D0-brane due to the contact interaction term in the DBI action) becomes proportional to
\beq
\de\left(k_2^0-k_1^0+\frac{P_2^2}{2M}-\frac{P_1^2}{2M}\right)\exp\left[-\frac{(P_2+k_2-P_1-k_1)^2d^2}{2}\right].
\label{resum}
\eeq
(Note that, because we have been considering normalized wavepacket states rather than momentum states, the momentum conservation $\de$-function does not appear in the amplitude in the $d\to\infty$ limit: rather, the amplitude is finite (in this limit) if momentum conservation is satisfied, and vanishes otherwise.)

The question is now how this amplitude should be expanded to match the structure of the IR-divergent CFT-based worldsheet perturbation theory. It turns out that the CFT amplitude is recovered (in the limit $d\to\infty$) by relating $d$ with the worldsheet cut-off $\eps$ via the identification 
\beq
d^2\sim g_{st}\alpha'\sqrt{|\log\eps|}.
\label{Dsquared}
\eeq
This relation is not obvious a priori and is imposed precisely because it makes the structure of the IR-divergent worldsheet perturbative expansion visible at the level of the DBI action. This IR-divergent structure is extremely unnatural from the standpoint of the DBI description, but it is forced upon us by the worldsheet CFT, where it is implemented by construction.

With the above identification, one can readily expand the amplitude (\ref{resum}) in powers of $g_{st}$ (or $1/M$) and observe the emergence of the IR-divergent structure reminiscent of the CFT. The lowest-order term is just
\be
\de\left(k_2^0-k_1^0\right),
\ee
which is obviously just one of the terms in the low-energy expansion of the worldsheet disk amplitude (keep in mind that we have restricted our analysis to the dilaton contact interaction term in the DBI Lagrangian). There are two corrections to the above $\de$-function arising at first order in $g_{st}$:
\beq
\left(\frac{P_2^2}{2M}-\frac{P_1^2}{2M}\right)\de'\left(k_2^0-k_1^0\right),
\label{one}
\eeq
which arises from expanding the energy conservation $\de$-function and
\beq
d^2(P_2+k_2-P_1-k_1)^2\de\left(k_2^0-k_1^0\right),
\label{two}
\eeq
which arises from expanding the exponential (keep in mind that $d^2\sim g_{st}$ if one is match the DBI and CFT descriptions).

If $P_1=P_2=0$, which is precisely the case of the standard D0-brane CFT, (\ref{one}) vanishes, whereas (\ref{two}) reproduces the IR-divergence in the modular integration of the annulus. The conclusion (reached here through an appeal to the DBI formalism) is then that the standard D0-brane CFT amplitudes without open strings describe\footnote{An alternative way to reproduce the infrared divergent CFT amplitudes within the DBI formalism would be to introduce a small ``mass'' $m$ for the ``field'' $X$ (that is, a shallow harmonic oscillator potential). Then, in the limit $m\to0$, the transition amplitudes for D0-brane in the ground state of the harmonic oscillator potential will reproduce the infrared divergent D0-brane CFT. Note that the ground state of the harmonic oscillator formally vanishes in the limit $m\to 0$, and it is in fact identical to the broad Gaussian wavepacket state employed in our derivations, provided that $d$ and $m$ are appropriately identified. We shall use this representation in appendix C.} a D0 brane whose initial and final states are Gaussian wavepackets centered around $P=0$, $X=0$, and the width should be identified with the CFT infrared cut-off as in (\ref{Dsquared}) and taken to infinity. (In the limit, the normalization factor of the wavepacket becomes infinite.)

Furthermore, through an appeal to the DBI formalism, we have given a complete resummation (\ref{resum}) of the leading infrared divergences in string amplitudes. The resulting picture is that the infrared divergences resum in a way that reconstructs the momentum conservation ``$\delta$-function''\footnote{In this paragraph, by ``$\delta$-function'' we mean a function which is 1 if its argument is 0, and 0 otherwise. It only differs from the ordinary $\de$-function by normalization, and makes an appearance in the amplitudes of this section due to the fact that we are working with normalized wavepacket states.}. Conversely, one can think of the infrared divergences in string perturbation theory as a result of attempting to expand in a Taylor series the (non-analytic) momentum conservation ``$\de$-function''. It is unfortunate that analogous resummations cannot be rigorously performed in full string theory (rather than the low-energy effective field theory) with present technology. Such resummations play an important role in the analysis of scattering of closed strings off D1-branes and the associated response phenomena \cite{local}. For recent investigations of infrared divergence resummations in the closed string sector, see \cite{sagnotti}.

\section{Normalization of the annular divergence}

In section 2, we have given a derivation of the functional form of the annulus divergence:
\beq
\left\langle V^{(1)}\cdots V^{(n)}\right>_{annulus}^{(div)}={\cal N}P^2\left\langle V^{(1)}\cdots V^{(n)}\right>_{D_2}\int\limits_0^1 dq\int\limits_{-\infty}^\infty d\omega\, q^{-1+\alpha'\omega^2}.
\label{Nrm}
\eeq
An integration over $q$ reveals the ``propagator'' of the translational mode:
\beq
\left\langle V^{(1)}\cdots V^{(n)}\right>_{annulus}^{(div)}=\frac{\cal N}{\alpha'}P^2\left\langle V^{(1)}\cdots V^{(n)}\right>_{D_2}\int\limits_{-\infty}^\infty \frac{d\omega}{\omega^2}.
\label{Nrmprpgtr}
\eeq
However, our considerations did not fix the overall normalization $\cal N$ in this expression.

It should be in principle possible to fix the coefficient $\cal N$ by
considerations within the worldsheet theory. This would require a careful
analysis of the unitarity constraints, which would determine the absolute normalizations of the vertex operators and the gluing parameter integration
measure in the plumbing fixture construction.

However, one can take advantage of the fact that the only ambiguity in (\ref{Nrm}) is a momentum-independent constant. Therefore, one can fix the 
value of this constant by considering the small momenta region. In this region,
the DBI set-up of appendix B would be applicable. Since unitarity
in field theory is automatic, determining normalization of the amplitudes
is much easier in the DBI set-up than in worldsheet theory. Furthermore,
since the coefficient $\cal N$ does not depend on which scattering process
is being considered, we can restrict ourselves to just one simple process,
for example, scattering of one dilaton off a D0-brane with the amplitude given by (\ref{matrix}):
\beq
\left<f\right|\int dt \exp\left[-i(k_2^i-k_1^i)X^i(t)\right]\exp\left[i(k_2^0-k_1^0)t\right]\left|i\right>.
\label{exp}
\eeq

To imitate within the DBI formalism the state of the D0-brane implicitly chosen by the D0-brane CFT, let us introduce a small ``mass'' $m$ for the ``field'' $X$:
\beq
S_{DBI}^{(m)}=M\int dt \left(\frac12 \left(\dot X\right)^2 - \frac{m^2}2X^2+ \frac12\left(\Phi(t,X^i(t))\right)^2+\cdots\right). 
\label{DBIm}
\eeq
This essentially generates a shallow harmonic oscillator potential $Mm^2X^2/2$ for the D0-brane. We can then choose $|i\rangle$ and $|f\rangle$ to be the ground state\footnote{It is in principle an assumption that the state of the D0-brane implicit in the (infrared-divergent) worldsheet CFT can be represented as the ground state of the harmonic oscillator in the mass-regulated DBI theory, and, in a rigorous treatment, it would need to be independently justified. We nevertheless feel that this assumption is plausible enough to be used, for practical purposes, in our computation of the coefficient of the annulus divergence.} of the harmonic oscillator. Then, the amplitude (\ref{exp}) can
be transformed\footnote{Note that, for any operator $A(t)$ in the interaction (or Heisenberg) picture, $\langle 0|A(t)|0\rangle=\langle 0|A(0)|0\rangle$.} as follows ($P^i=k_1^i-k_2^i$):
\beq
\ber{l}
\dsty\left<0\right|\int dt \exp\left[-i(k_2^i-k_1^i)X^i(t)\right]\exp\left[i(k_2^0-k_1^0)t\right]\left|0\right>\vspace{3mm}\\
\dsty\hspace{3cm}\sim\de(k_2^0-k_1^0)\left\{1-\frac{P^iP^j}{2}\left<0|X^i(0)X^j(0)|0\right>+\cdots\right\}.
\eer
\label{expexpand}
\eeq
(expanding the exponential in Taylor series generates Feynman graphs for the field $X$; the first term in the brackets is the leading order contribution, the second term
is the lowest order loop correction). To make contact with (\ref{Nrmprpgtr}), we shall express $\left<0|X^i(0)X^j(0)|0\right>$ through the propagator of the $X$-field:
\beq
\left<0|X^i(0)X^j(0)|0\right>=\frac{\de^{ij}}{2\pi M}\int\frac{d\omega}{\omega^2+m^2}
\eeq
Then, the ratio between the tree-level (disk) and the one-loop
(annulus) contributions in (\ref{expexpand}) is given by
\begin{eqnarray}
-\frac{P^{2}}{4\pi M}\int\frac{d\omega}{\omega^2+m^2}.
\label{loop-DBI}
\end{eqnarray}

In order to compare the DBI expression (\ref{loop-DBI}) and the worldsheet expression (\ref{Nrmprpgtr}), we note that, in both formulas,
$\omega$ denotes the same physical quantity: it is the energy of the off-shell
massless open string state propagating in the loop. We can then introduce a common infrared regulator
in (\ref{loop-DBI}) and (\ref{Nrmprpgtr}) by removing energies less than $\omega_{min}$ and sending $m$ to 0. This yields
\beq
-\frac{P^{2}}{4\pi M}\int\limits_{|\omega|>\omega_{min}}\frac{d\omega}{\omega^2}
\label{loop-DBI'}
\eeq
for (\ref{loop-DBI}), and
\beq
\frac{\cal N}{\alpha'}P^2\int\limits_{|\omega|>\omega_{min}} \frac{d\omega}{\omega^2}
\label{Nrmprpgtr'}
\eeq
for (\ref{Nrmprpgtr}). Comparing these two expressions, one can deduce
the value of the normalization coefficient $\cal N$:
\beq
{\cal N}=-\frac{\alpha'}{4\pi M}.
\label{normalization}
\eeq

Now we can substitute this value of $\cal N$ into (\ref{Nrm}) and obtain the
expression for the annulus divergence with the correct normalization. For comparison with worldsheet derivations, it is convenient to cut off the
small values of the gluing parameter $q$ in order to regularize the
infrared divergence, as was done in section 2. Then,
\beq
\ber{l}
\dsty\left\langle V^{(1)}\cdots V^{(n)}\right>_{annulus}^{(div)}=-\frac{\alpha'P^2}{4\pi M}\left< V^{(1)}\cdots V^{(n)}\right>_{D_2}\int\limits_\eps^1 dq\int\limits_{-\infty}^\infty d\omega\, q^{-1+\alpha'\omega^2}\vspace{3mm}\\
\dsty\hspace{3cm}=-\frac{P^2}{2M}\,\sqrt\frac{\alpha'}{\pi}\,\left< V^{(1)}\cdots V^{(n)}\right>_{D_2}\sqrt{|\log\eps|}.
\eer
\label{normalized}
\eeq

\end{document}